\documentclass[letterpaper,longauthor,trackchanges,twocolumn,onecolappendix,sort&compress]{aastex62}

\usepackage{hyperref}
\usepackage{multirow}
\usepackage[english]{babel}
\usepackage{amsmath}
\usepackage{amssymb}
\usepackage{graphicx}

\def\apj{AstroPhysical Journal}             
\def\apjl{Astrophysical Journal, Letters}                
\def\apjs{Astrophysical Journal, Supplements}  \def\nat{Nature}               %
\def\mnras{Monthly Notices of the Royal Astronomical Society}
\def\prl{Physics Review Letters}
\def\procspie{Proceedings of the SPIE}

\newcommand{\dtone}{\ensuremath{\Delta T_1}}
\newcommand{\dttwo}{\ensuremath{\Delta T_2}}

\begin{document}

\title{{\em Presto-Color:} A Photometric Survey Cadence for Explosive Physics \& Fast Transients}

\correspondingauthor{Federica B. Bianco}
\email{fbianco@udel.edu}

\author[0000-0003-1953-8727]{Federica B. Bianco}
\affil{{Department of Physics and Astronomy, University of Delaware, Newark, DE, 19716, USA
2}}
\affil{{Joseph R. Biden, Jr. School of Public Policy and Administration, University of Delaware, Newark, DE, 19716, USA
2}}
\affil{{Data Science Institute, University of Delaware, Newark, DE, 19716, USA
2}}
\affil{{Center for Urban Science and Progress, New York University, 370 Jay St,
Brooklyn, NY 11201, USA}}
\author{Maria R. Drout}
\affil{{Department of Astronomy and Astrophysics, University of Toronto, 50 St. George Street, Toronto, Ontario, M5S 3H4 Canada}}
\affil{The Observatories of the Carnegie Institution for Science, 813 Santa Barbara St., Pasadena, CA 91101, USA}
\author{Melissa L. Graham}
\affil{Department of Astronomy, University of Washington, Box 351580, U.W., Seattle, WA 98195, USA}
\author{Tyler A. Pritchard}
\affil{Center for Cosmology and Particle Physics, New York University, 370 Jay St,
Brooklyn, NY 11201, USA}
\author{Rahul Biswas}
\affil{The Oskar Klein Centre for CosmoParticle Physics, Department of Physics, Stockholm University, AlbaNova, Stockholm SE-1069}
\author[0000-0001-6022-0484]{Gautham Narayan}
\affil{Space Telescope Science Institute, 3700 San Martin Dr, Baltimore, MD 21218, USA}
\author[0000-0002-8977-1498]{Igor Andreoni} 
\affil{Division of Physics, Mathematics and Astronomy, California Institute of Technology, 1200 E California Blvd, Pasadena, CA 91125, USA}
\author[0000-0002-2478-6939]{Philip S. Cowperthwaite}
\affil{The Observatories of the Carnegie Institution for Science, 813 Santa Barbara St., Pasadena, CA 91101, USA}
\author{Tiago Ribeiro}
\affil{LSST, 933 N. Cherry Ave., Tucson, AZ 85721, USA}

\collaboration{with the support of the LSST Transient and Variable Stars Collaboration}


 \begin{abstract}
 
We identify minimal observing cadence requirements that enable photometric astronomical surveys to detect and recognize fast and explosive transients and fast transient features. Observations in two different filters within a short time window (e.g., $g$-and-$i$, or $r$-and-$z$, within $<0.5$ hr) and a repeat of one of those filters with a longer time window (e.g., $>1.5$ hr) are desirable for this purpose. Such an observing strategy delivers both the color and light curve evolution of transients on the same night.     This allows the identification and initial characterization of fast transient---or fast features of longer timescale transients---such as rapidly declining supernovae, kilonovae, and the signatures of SN ejecta interacting with binary companion stars or circumstellar material. Some of these extragalactic transients are intrinsically rare and generally all hard to find, thus upcoming surveys like the Large Synoptic Survey Telescope (LSST) could dramatically improve our understanding of their origin and properties. We colloquially refer to such a strategy implementation for the LSST as the {\em Presto-Color} strategy (rapid-color). This cadence's minimal requirements allow for overall optimization of a survey for other science goals. Published version DOI: \url{https://doi.org/10.1088/1538-3873/ab121a}.
\end{abstract}


%
\section{Introduction}\label{sec:intro}

The advent of wide-field time domain surveys has revolutionized the field of transient astrophysics. Coverage on short timescales in the optical/NIR bands, in particular, has facilitated rapid strides in our understanding of both supernova (SN) explosions and peculiar transients. This work focuses on short-lived transients, whose light curves evolution happens in $\lesssim$10 days, as well as slower transients (evolving in months time scales) that show fast evolving features at some phases.

Rapidly evolving transients are generally poorly understood, and may be associated with a variety of phenomena, including accretion-induced white dwarf collapse \citep{Metzger2009}, underluminous and fallback SN~\citep{Moriya2010}, ultra-stripped SN \citep{Drout2013,Kasliwal2010,Tauris2015,De2018}, compact-object mergers \citep{Kasen2015,Metzger2010}, orphan gamma-ray burst afterglows \citep{Totani2002}, and common-envelope ejections \citep{Blagorodnova2017}. ``Infant'' supernovae (SNe)---hours to days after explosion---evolve quickly and their observations provide vital constraints on their explosion mechanisms and progenitor systems \citep{Nakar2010,Rabinak2011,Nugent2011}, potential non-spherical behavior \citep{Matzner2013,Salbi2014}, and shock collision with a binary companion \citep{Kasen2010}. 


Despite progress, the detection rate for both rapid transients and rapidly evolving \emph{phases} in SN explosions has remained low due to a combination of survey efficiency and intrinsic event rates. The volume surveyed by the new and upcoming surveys Zwicky Transient Factory (ZTF; \citealt{ztf}) and the Large Synoptic Survey Telescope (LSST; \citealt{lsst}\footnote{as accessed in its \url{arxiv.org} version, which is a living document, on January 30, 2019}) bring the promise of detecting many more intrinsically rare events. However, recognizing and using these events to probe the science questions described herein (\autoref{sec:transients}) requires adequate time- and filter-sampling of relatively short-lived features. 

In this work, we explore the minimal cadence requirements that allow a multi-band imaging survey to effectively \emph{recognize} young and rapidly evolving transients. We focus on an implementation for the LSST main survey (also known as Wide-Fast-Deep, WFD). The LSST main survey is designed with four science goals in mind: understanding dark matter and dark energy, cataloging the orbits of millions of moving objects in the solar system, understanding the structure of the Milky Way via resolved stellar population, and exploring the transient sky \citep{lsst}. To deliver on these science goals, the LSST is planning a 6-filter (\emph{ugrizy}), $\sim18,000$~square degree survey of the southern hemisphere to a single-image (coadded) depth of $r\gtrsim24.5$ mag ($27.5$ mag) for a signal-to-noise ratio of $5$ for point sources, with subsecond spatial resolution \citep{lsst}\footnote{For more details, see also the LSST Science Requirements Document \url{https://ls.st/srd}.}. However, the LSST main survey's baseline cadence results in a typical intra-night revisit rate of $\sim$30 minutes \citep{COSEP}, which is too rapid to recognize effectively fast transients and fast transient features from ``normal'' transients. Furthermore, the current baseline planned cadence repeats observations in the same filter with a median gap of 2 weeks: too long to capture rapidly evolving transients with enough epochs for characterization, and in extreme, but also extremely interesting cases, \emph{e.g.} kilonovae (KNe), to even capture them with anything more than a single epoch. We are most interested in optimizing the cadence of LSST because, compared to other ongoing and upcoming all sky transient surveys, LSST will have the greatest depth. This provides better uncertainties on the flux compared to other surveys, enabling a more accurate color or  light curve evolution determination from a small number of observations, and could be critical in the determination of these quantities from a relatively small number of observations. Second, the higher depth allows LSST to probe a larger volume for the same sky area, which is better for building a sample of intrinsically rare transients and features --- but a key constraint is the potential availability of spectroscopic follow-up facilities for the fainter targets.

The optimization of the LSST survey strategy is, at the time of writing, open to community input, and the LSST Project Science Team issued a call for white papers suggesting LSST cadences details in 2018 November, which prompted this work and many other strategy implementation ideas\footnote{https://www.lsst.org/submitted-whitepaper-2018}. 

We explored the \emph{minimal} requirements necessary to achieve the goal of detecting and recognizing fast transients and fast transient features, and design a strategy for the LSST main survey (WFD) that accommodates the core LSST science goals (and other science goals identified by the scientific community) but which, unlike many other WFD proposals (including the current baseline strategy\footnote{the {\tt baseline2018a} simulation described in https://ls.st/Document-28453}), allows for the identification of young and rapidly evolving transients from the millions of LSST alerts every night. This cadence will allow for the prompt triggering of external follow-up resources (e.g., spectroscopy, non-optical facilities) that are necessary to confirm the discoveries and study the temperature, composition, and explosion mechanisms of explosive phenomena.

We argue that three observations per night are sufficient to disambiguate fast transients and trigger prompt follow-up, as long as the three images are collected in two filters with appropriate time gaps between the filters in the sequence. We find that this cadence should put minimal strain on an observing strategy, and still allow for further strategy details to be optimized for different science cases (e.g. covering large areas of the sky within a night, or the specific pointing sequence, or obtaining long observation sequences in a single filter). Specifically, the strategy we envision has two requirements: 
\begin{enumerate}
    \item Observations in {\bf two filters} obtained in quick succession so that the transient's {\bf color} can be measured. The color is critical to both allow us to distinguish different classes of transients and as a probe of the physics operating during rapid transient evolution phases.
    \item A {\bf same-filter} revisit separated by hours (before or after the filter pair) so that the light curve {\bf evolution (slope)} can be constrained and distinguished from slower-evolving transients.
\end{enumerate}
Since this cadence is designed to return both a transient's color and rate of brightness change in a single night, we call it the {\em Presto-Color} (rapid-color) cadence. 
The exact form of the {\em Presto-Color} cadence is provided with more detail in \autoref{sec:implementation}.


In order to demonstrate the ability of the {\em Presto-Color} survey cadence to achieve our goals, we have selected four exemplary types of extragalactic fast transients, and fast features of longer duration transients, for our discussion in \autoref{sec:transients}. Representative light curves for each type of transient are shown in \autoref{fig:examples}, and graphical representations of where they separate from normal SN in color and intra-night rate-of-change are shown in \autoref{fig:phasespace} and \autoref{fig:classifier}.  

\section{Scientific Motivation}\label{sec:transients}

\begin{center}
\begin{figure*}[!t]
\includegraphics[height=4.4cm]{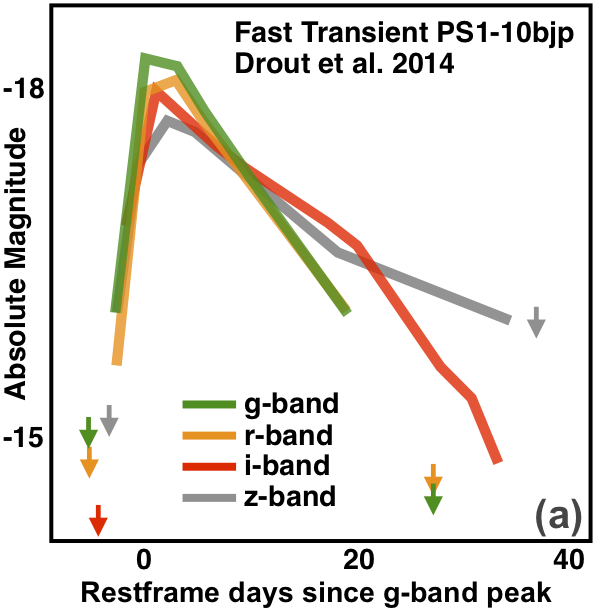}
\includegraphics[height=4.4cm]{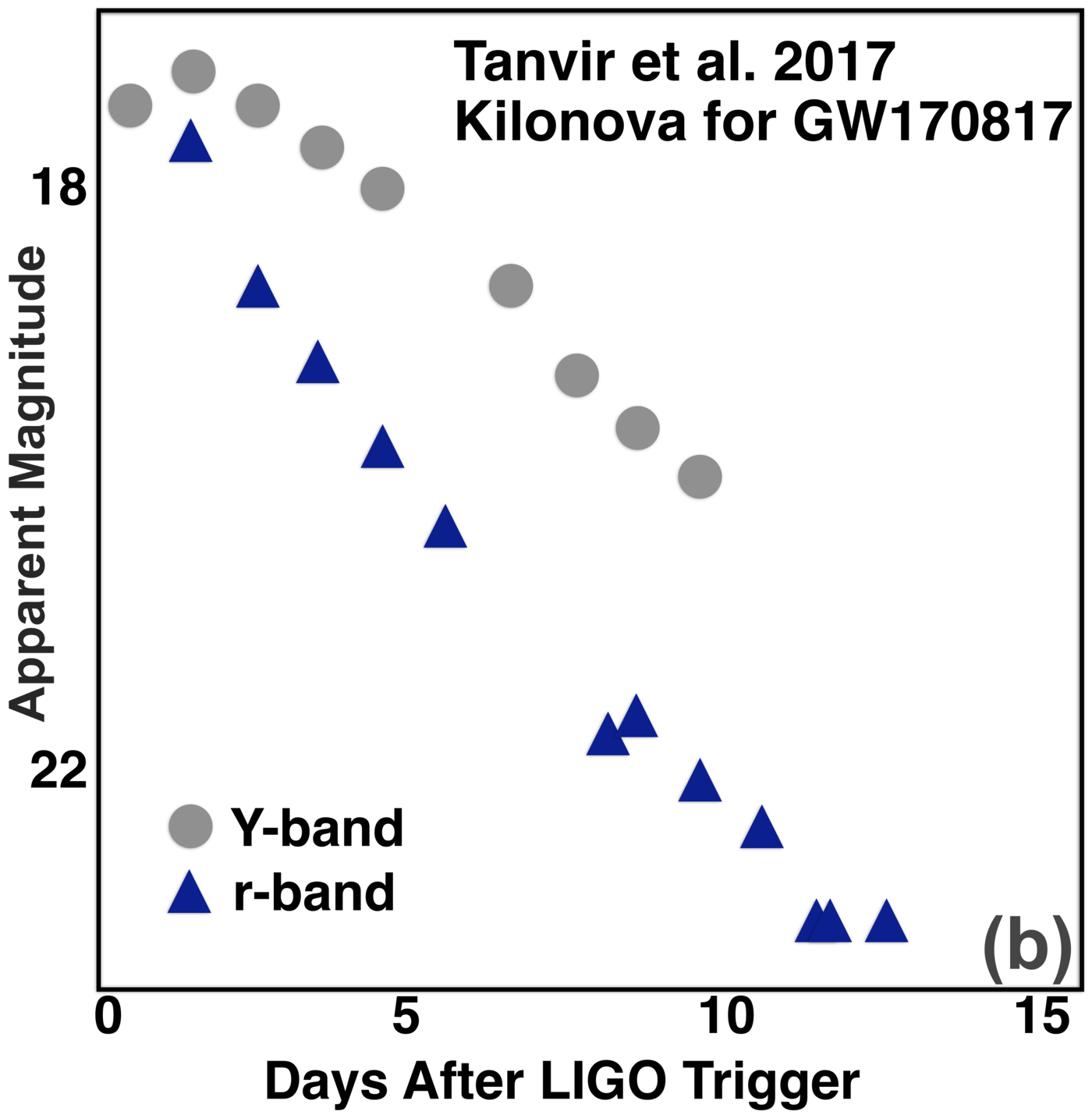}
\includegraphics[height=4.4cm]{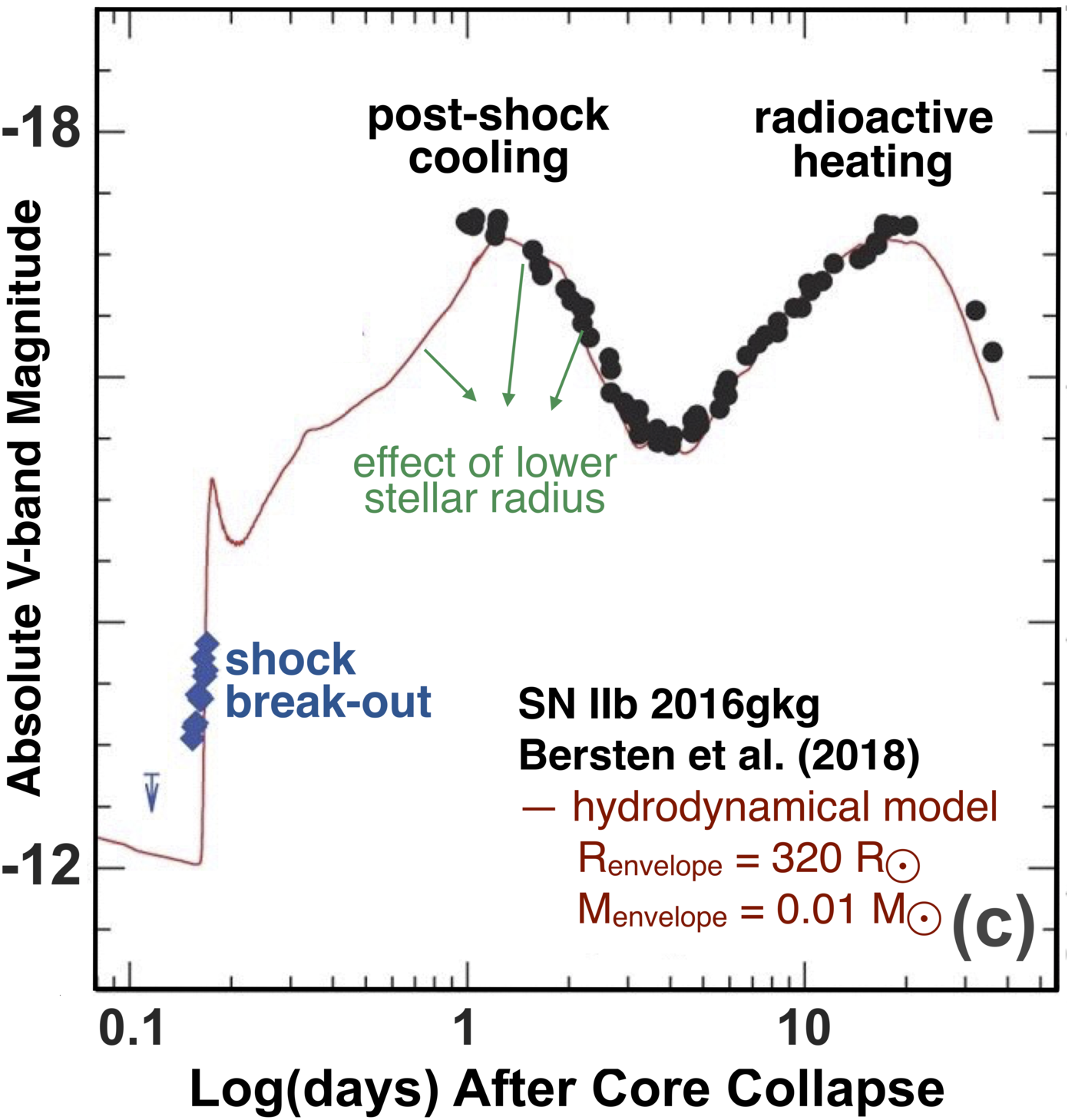}
\includegraphics[height=4.4cm]{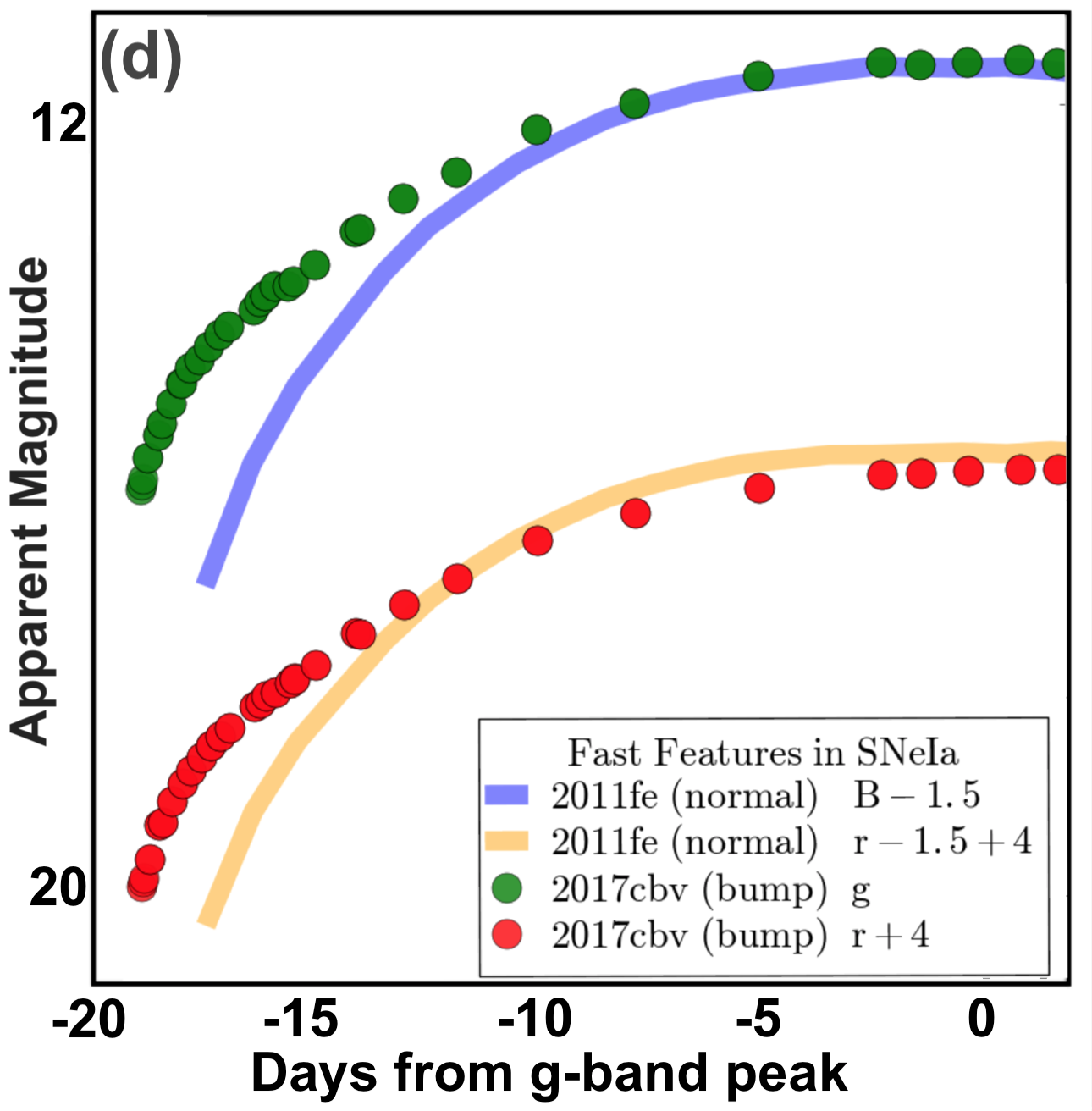}
\caption{{Light curves of  exemplary fast transients and fast features. From left to right: (\emph{a}) fast transient PS1-10bjp \citep{Drout2014}; (\emph{b}) the kilonova associated to GW170817 \citep{Tanvir2017}; (\emph{c}) the shock breakout model fits for SN\,IIb 2016gkg's stellar radius \citep{Bersten2018}; and (\emph{d})  Type Ia SN 2017cbv's ``blue bump'' compared to the ``normal'' Type Ia SN 2011fe~\citep{Graham2015, Hosseinzadeh2017}.} In some cases, these plots have been adapted from their original form for clarity in the context of this paper.}\label{fig:examples}
\end{figure*}
\end{center}

The main science cases that motivate the design of the {\em Presto-Color} cadence are discussed below, and simplified light curves are shown in \autoref{fig:examples}.

\subsection{The Nature of Rapidly Evolving Luminous Transients} 

``Rapidly evolving transients'' are defined as extragalactic events that reach SN luminosity but have timescales an order of magnitude faster. To date, only a small number have been identified, but recent sample studies \citep{Drout2014} have shown that they are not \emph{intrinsically} rare: few have been detected simply because current surveys are not designed to be efficient at short timescales. They are a significant fraction ($\sim$5\%--10\%) of the core-collapse events, which we must understand to have a complete picture of massive star death. Known events have rise times spanning 1--3 days and blue colors at maximum \citep{Drout2014,Pirsiainen2018,Rest2018}; an example of a fast transient light curve is shown in \autoref{fig:examples} panel (a). While their true nature is unknown, leading theoretical models include black hole formation in failed SN and the birth of binary neutron star systems, with recent observations of AT2018cow showing evidence for a central engine \citep{Kashiyama2015,Margutti2018,Prentice2018}. 
It is clear that larger samples, as well as more detailed and complete observations of individual event are required to understand their true nature and diversity.

\subsection{KNe and the Origin of Heavy Elements} 

KNe are powered by the radioactive decay of \emph{r}-process nuclei synthesized in the ejecta of neutron star mergers \citep{Li1998,MetzgerKN}.  These objects have been described by theories and models, and candidate KNe have been observed in conjunction with Gamma Ray Bursts \citep{Berger13, Tanvir13, Jin15}. The first detection of an optical counterpart associated with a gravitational wave (GW) event came in 2017 in a search triggered by the \emph{Advanced LIGO} \citep{Aasi15} and \emph{Advanced Virgo} \citep{Acernese15} detection of 
GW170817 \citep{Abbott2017}. Observations of the KN associated with GW170817 revealed thermal emission that rose in $<$1 day and cooled from a temperature of $>$10,000 K to 3,000 K over 5 days \citep{Drout2017}. The initially blue optical light faded at a rate of $>$1 mag $\mathrm{day}^{-1}$, and was followed by a longer-lived red transient consistent with the production of a significant quantity of \emph{r}-process elements of \emph{multiple} compositions \citep[][and references therein]{Cowp+17,Drout2017,Kasliwal2017,Smartt2017,Tanvir2017,Villar2017}. An example of this $\sim$10 day long red light curve is shown in \autoref{fig:examples} panel (b). Additional examples --- with or without associated LIGO triggers --- are required to ascertain whether GW170817 was typical. Sample studies will provide constraints on the ejecta composition, mass, and velocity that strongly influences the resulting \emph{color}, \emph{magnitude}, and \emph{timescale} of the emission. Once the effect of the ejecta on the color is understood, the frequency of the early blue emission will set critical constraints on the ratio of light and heavy elements formed, and the total contribution of NS mergers to cosmic nucleosynthesis \citep{Metzger2018,Piro2018,Rosswog2018}.

Unfortunately, the current baseline cadence planned for the LSST WFD Survey repeats observations in each filter with a median gap of 2 weeks. This is longer than the anticipated time scale of KNe, as well as many other putative systems that produce fast transients. The call for LSST white papers recognized that this need is largely unmet, and several cadence proposals are aimed at improving LSST's ability to probe short ($< 1$~week) duration time-domain phenomena. Notably, responding to the same call for white papers, \citet{Andreoni2018} advocates for an LSST ``rolling'' cadence with colors for the purpose of increasing the chance detection rate of KNe. This strategy is similar to ours advocating for 2-filter (\emph{g} and \emph{i}) observations on consecutive days. 


\subsection{Progenitors and Pre-explosion Mass Loss of Core-collapse SN} 

Early observations of core-collapse SN (CCSN) provide critical constraints on the progenitor radius and envelope structure through the detection of either shock breakout ($\sim$1 day) or cooling envelope ($\sim$1-5 day) emission \citep{Modjaz2009,Nakar2010,Arcavi2011,Bersten2018}. An example of a CCSN that exhibited both types of fast features, and a hydrodynamical model fit to the data, is shown in \autoref{fig:examples} panel (c). Indeed, there has been growing evidence that many CCSN either explode in ``non-standard'' evolutionary states or undergo enhanced pre-SN mass-loss and outbursts in their terminal years \citep{Nakar2014,Khazov2016}. Theoretical studies have pointed to a range of potential explanations to accommodate the observations, such as pulsation-driven superwinds \citep{Yoon2010}, wave heating outbursts \citep{Fuller2017}, and inflated progenitor envelope \citep{Grafener2012}. However, the nature of this mass loss and the types of SN experiencing it remain uncertain.

\subsection{Progenitors and Explosion Mechanisms of Thermonuclear SN}

SN Ia result from the thermonuclear disruption of a carbon-oxygen white dwarf star which has either accreted mass from or merged with a binary companion \citep{Hillebrandt2000}. However, questions remain regarding the nature of the companion star (e.g., a red giant, main sequence, or another white dwarf star). Answering these questions are not only important for a complete and accurate picture of stellar evolution and death, but also for understanding how the diversity of progenitor systems and differences in the SN evolutionary pathways over cosmic time may induce systematic differences in standardization, and thus systematic errors in precision cosmological inference (e.g., \citealt{2011PhDT........37D,2013PhDT.......326H}).

Observable signatures of a stellar companion star in the progenitor system of a SN\,Ia is, for the above reasons, an active pursuit of the SN community. For example, \citet{Kasen2010} predicted that a red giant or main sequence companion star could shock the SN\,Ia ejecta at early times, and cause an observable ``blue bump" in the first few days of a SN\,Ia light curve. In this model the distance and size of the companion star relative to the white dwarf drives the color and duration of this fast blue light curve feature, while the viewing angle of the observer relative affects the intensity of the observed light curve ``bump.'' Sample studies of type Ia SN lead to constraints on the progenitor fractions \citep{Bianco11, Hayden2010}. Singular examples of ``blue bumps" have been observed: for example, observations within $\sim$1 day of explosion revealed a rapidly-rising blue ``bump'' for SN\,Ia 2017cbv (panel (d) of \autoref{fig:examples}), which has been interpreted by some as a collision with a non-degenerate companion star \citep{Hosseinzadeh2017}.

\begin{figure*}[!t]
\begin{center}
\includegraphics[width=0.9\textwidth]{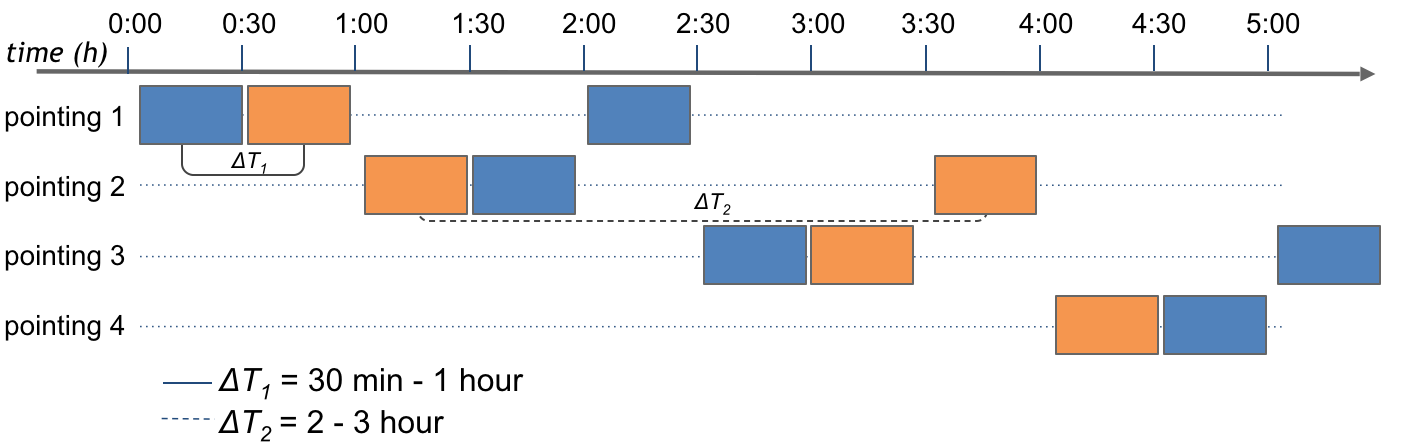}
\caption{A schematic example of the {\em Presto-Color} cadence with two alternating filters covering four regions of sky over 5.5 hours to obtain three observations per region with appropriate time gaps to measure light curve color and shape. In this implementation the time gaps are: $\mathrm{30~min} \lesssim \dtone\lesssim \mathrm{1~hour}$ and $\mathrm{2~hour} \lesssim \dttwo\lesssim \mathrm{3~hour}$ . Preliminary results for such an implementation of {\em Presto-Color} for LSST (performed with {\tt OpSim}, see \autoref{sec:implementation} and Appendix) indicate a $\sim1\%$ efficiency loss due to the increased number of filter changes.}\label{fig:implementation}
\end{center}
\end{figure*}

However, larger sample studies are required to resolve the degeneracy between the nature of the companion and the geometric alignment, and to furthermore help distinguish between the effects of a companion star and the influence of the explosion itself on the early light curve. For example, preliminary population studies suggest the possibility of an as-yet-unexplained red/blue color dichotomy in early ($<$ 5 days) rapidly rising light curves of SN\,Ia \citep{Stritzinger2018}, and implications for outwardly mixed radioactive material are predicted by the double detonation explosion model \citep{Piro2016,Polin2018}. The latter physical quality may also be related to dust formation models from SNe, and also important for sub-Chandrasekhar mass models for SN\,Ia progenitors \citep{Polin2018}. The {\em Presto-Color} cadence for LSST will not only help to identify such fast features early enough for follow-up, but lead to a significantly higher fraction of LSST's discovered SN\,Ia having the requisite early time sampling for large statistical analyses.

\subsection{Additional Science Cases}\label{sec:additional}

While we have focused on extragalactic fast transients here, a cadence that allows measurement of both color and rate-of-change on the timescale of $\sim$hours will have general applicability across many areas --- from variable stars and microlensing to characterization of solar system objects. Requests for multiple observations in the same and in different filters, in fact, were made in several other white papers submitted to the LSST Project in November including \citet{streetWP}, which focuses on microlensing, and  \citet{Bricman2018} and \citet{Gezari2018}, both of which focus on tidal disruption events.

\section{Implementation}\label{sec:implementation}

The {\em Presto-Color} cadence aims to obtain both the color and light curve shape of a transient in a single night and it is originally designed as an LSST WFD strategy. It requires observations in two different filters, $f_1$ and $f_2$, within a short interval of time \dtone, and then to return to the same field with either of those filters at a later time \dttwo. The minimal technical constraints in this observing strategy are:
\begin{enumerate}
    \item {\tt max(\dtone)}, an upper limit on the time between the two visits that provide color,
    \item {\tt min(\dttwo)}, a lower limit on the time between the two visits that provide light curve shape, and
    \item The filter pair $f_1$ and $f_2$.
\end{enumerate}

As we envision the implementation of {\em Presto-Color} for a synoptic survey, we have to keep in mind efficiency is generally a priority for such surveys, and for LSST in particular: the implementation of {\em Presto-Color} must not interfere with the achievement of the four main science goals described in \autoref{sec:intro}. We tested a simple implementation of a strategy that delivers on our requirements by using the LSST {\tt OpSim}\footnote{https://www.lsst.org/node/656} software \citep{opsim}: an application that simulates the field selection and image acquisition process of the LSST survey over the lifetime of the survey, balancing  strategic requirements on different time scales (e.g. minimum number of observations per filter per field over the survey lifetime \emph{vs} target filter gaps and filter sequences), simulating realistic weather and seeing conditions, scheduled engineering downtime, and current telescope and camera parameters (a more detailed discussion of our simulations is presented in the Appendix). As illustrated in \autoref{fig:implementation}, the {\em Presto-Color} cadence can be implemented by alternating pairs of visits on a field and single visits on the previous field, thus minimizing slew time. The single visits separated by \dttwo\ could alternate between the two filters, thereby reducing the number of filter changes (which incurs a $2$ minute overhead with LSST). In the example implementation of \autoref{fig:implementation}, half of the transients would have their light curve slope measured in ``blue" filter, the other half in ``orange" filter, and all would have a data point in ``blue-minus-orange" or ``orange-minus-blue'' color. 

The main science goals of {\em Presto-Color} might be reachable if the proposed cadence is limited to extra-galactic LSST fields, although we noted in \autoref{sec:additional} that some galactic transients (e.g., microlensing) would also benefit from it. Obtaining two observations within a short time interval, comparable to our \dtone, is necessary to distinguish moving objects (\emph{i.e.} Solar System objects) from transients and variables, so a repeat observation within a short interval of time is built in almost all LSST cadence implementations. However, collecting these observations in two different filters $f_1$ and $f_2$  limits the detectability of Solar System objects to the sensitivity of the shallowest filter in the $f_1-f_2$ pair. The {\em Presto-Color} cadence could be avoided in regions of the extended ecliptic plane (i.e., the region of interest for solar system studies). Essentially, given the incredible increase in the LSST's accessible volume compared to past surveys, implementing the {\em Presto-Color} cadence in just some sky regions, some of the time, should yield more (and better) observations of fast transients and fast features than ever before.

\section{Recognizing fast transients and fast features}

We explore our ability to recognize fast transients and fast features in transients, distinguishing them from ``normal'' transients, assuming an observing strategy that collects three images within a single observing night. We parameterize transients in a phase space of observed color versus observed intra-night magnitude-change to determine which region of this space each transient occupies.


\subsection{Sample Light Curves}

We first compile a sample of light curves for both fast transients and ``normal'' supernova from the literature, which will be used to test the effectiveness of different observing strategies in distinguishing targets of interest. 

For our sample of fast transients/features, we include example light curves for each of the classes of events described in \autoref{sec:transients}.
Our sample of rapidly evolving transients includes the gold sample of events from \cite{Drout2014} as well as the rapidly declining transients SN\,2005ek \citep{Drout2013} and AT2018cow \citep{Margutti2018}. For KNe, we consider the fiducial case of the best fit models to the emission of GW170817 from \citet{Villar2017}. We include models both with and without the early blue component (labeled as 'Blue Kilonova' and 'Red Kilonova'). For ``infant'' SNe we consider observations of supernova obtained within the first 5 days after explosion. We include both empirical data from the core-collapse SN\,2016gkg (supplemented by a hydrodynamical model from \citealt{Piro2016} to obtain finer time steps) and the Type Ia SN2011fe and SN2017cbv \citep{Graham2015,Hosseinzadeh2017} as well as simulated observations of SNe Ia with and without binary companions based on the models of \citet{Kasen2010}.

\begin{figure}[!t]
\begin{center}
\includegraphics[width=0.5\textwidth]{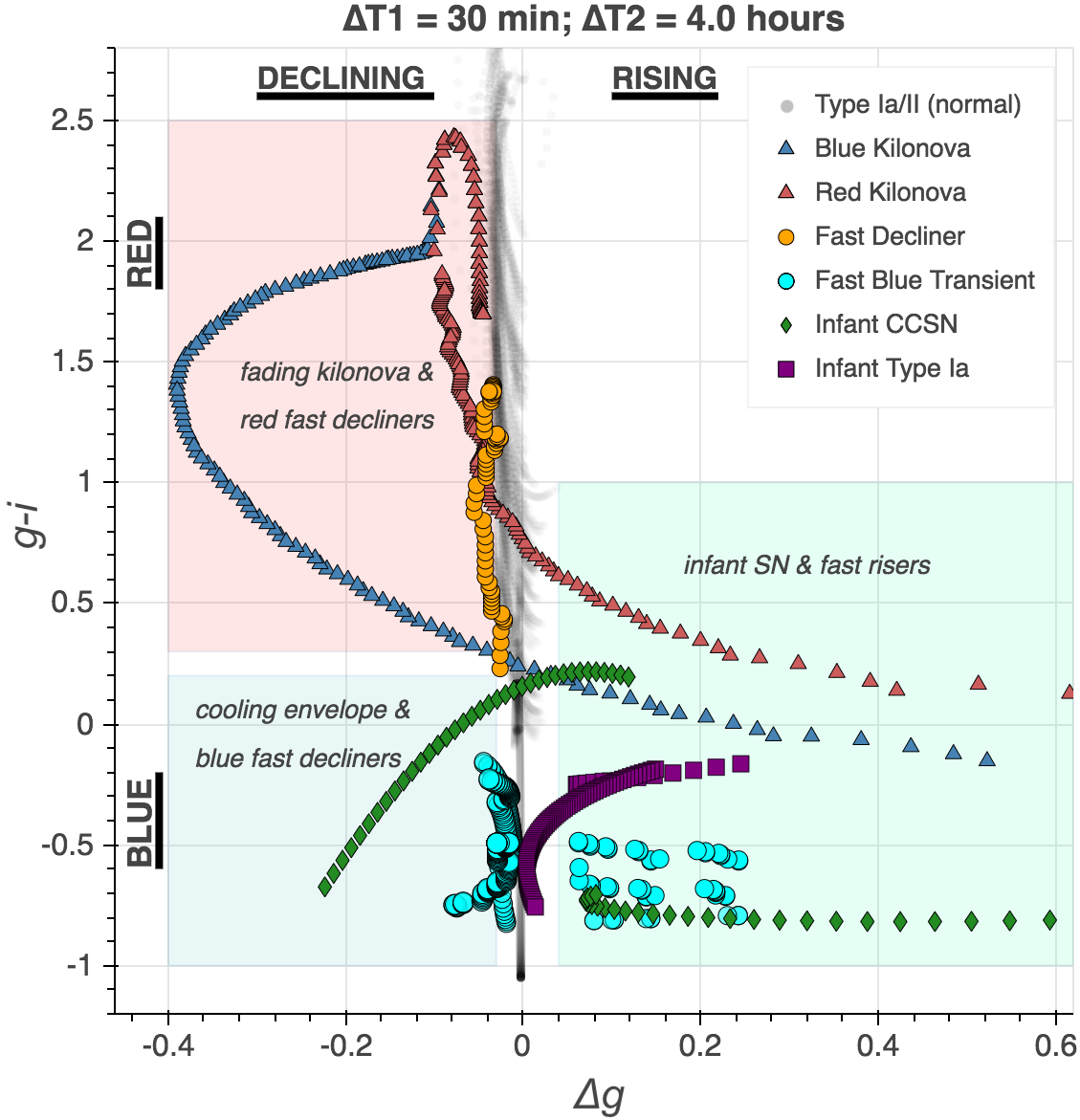}
\caption{Fiducial phase space plot showing separation between classes of transients in observed color and intra-night magnitude change. We show the location of transients for observations obtained in $g$-and-$i$ band filters within 30 minutes (\dtone) and a second $g-$band observation obtained 4 hr (\dttwo) after the first. All fast transients/features are shown in colors, SNe Ia, as representatives of ``normal transients,'' are plotted in gray, clustering around 0 on the $x$-axis due to their slower evolution (see legend and text for details). Each transient is shown at a range of epochs in its evolution. Rising light curves correspond to positive magnitude changes (positive $x$-axis).  }\label{fig:phasespace}
\end{center}
\end{figure}

\begin{figure*}[!t]
\begin{center}
\includegraphics[width=0.49\textwidth]{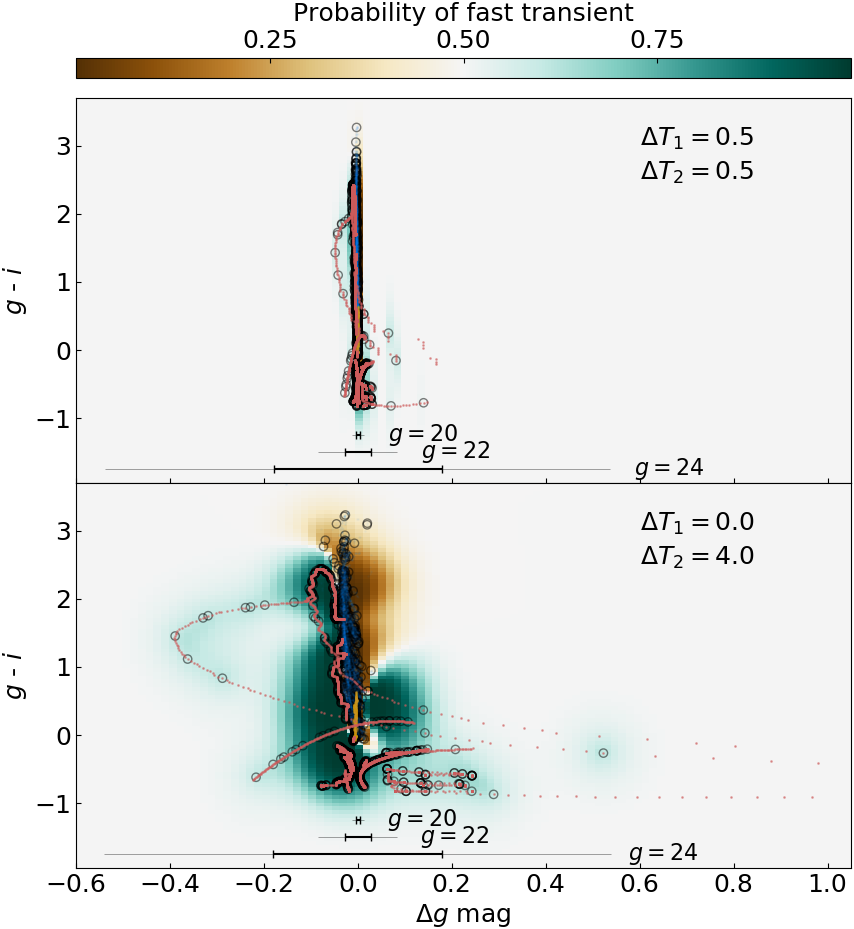}
\includegraphics[width=0.49\textwidth]{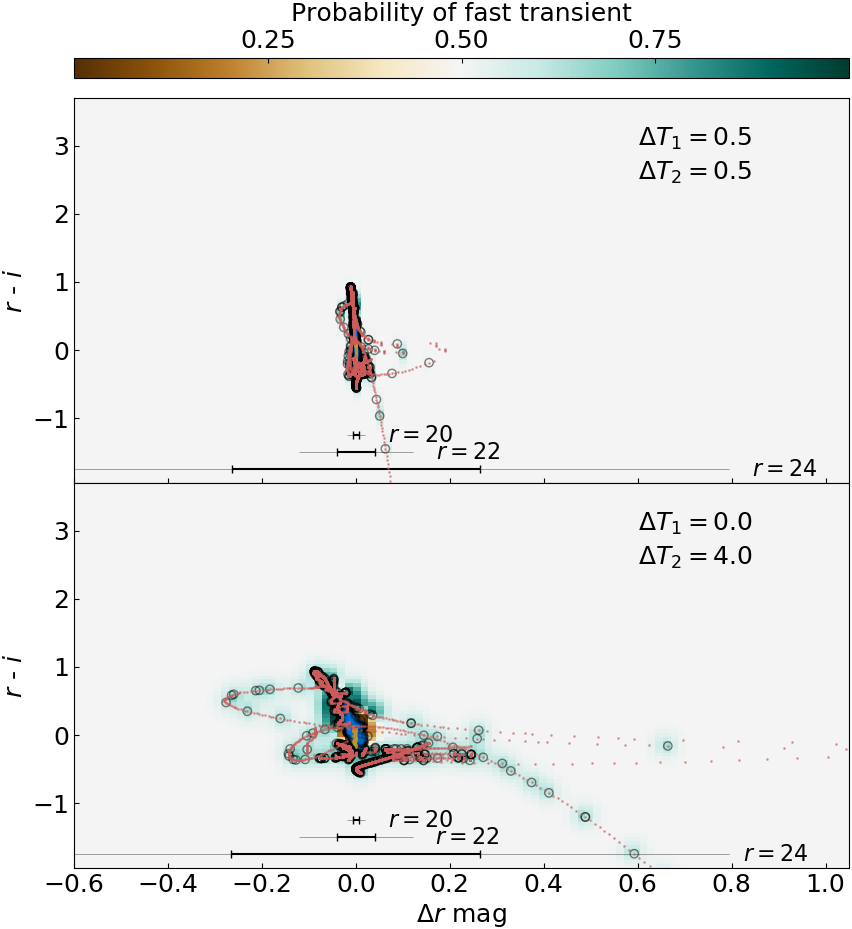}
\caption{{The result of a binary classification (``normal'' \emph{vs.} ``fast'') performed with a Gaussian Processes (GP)  Probabilistic Classifier for two fiducial strategies: (\dtone,\dttwo) = (0.5 hr, 0.5 hr), shown at the \emph{top}, and (\dtone,\dttwo) = (0, 4 hr), at the \emph{bottom}. The \emph{left} panel shows the result for observations in $g$-and-$i$, the \emph{right} panel in $r$-and-$i$, with the $g$ and $r$ filter repeated twice, respectively,  leading to a light curve shape constraint $\Delta g$ and $\Delta r$. The red dots indicate transients of interest (same sample as the colored data points in \autoref{fig:phasespace}) and blue and gold dots denote SN Ia and II populations respectively. Gray circles indicate the points included in the training set. The heat map in the background represents the probability of a transient being ``not-normal'' from three observations collected according to the specific implementation of our \emph{Presto-Color}  strategy. The majority of the phase space has a probability $p~\sim~0.5$ because it is far from any known transients: the region densely occupied by known transients is relatively compact, except when observed in $g$-and-$i$ at (\dtone,\dttwo) = (0, 4 hr). The error bars at the bottom indicate the photometric uncertainty in $\Delta mag$ based on the typical rate of change for a normal SN\,Ia (0.3 mag/day) for a transient initially observed by LSST at mag~20, 22, and 24 in the $g$ (\emph{left}) or $r$ (\emph{right}) band: thick lines represent a $1\sigma$, thin lines a $3\sigma$ uncertainty. Transients at a horizontal separation smaller than an error bar have indistinguishable light curve evolution within that confidence level. When observed with gaps (\dtone,\dttwo) = (0.5 hr, 0.5 hr) nearly all transients  $g/r\gtrsim22$ at nearly all epochs are indistinguishable at a $1\sigma$ level in both filter combinations. Even at $g/r\sim20$ transients are indistinguishable at a high ($3\sigma$) confidence.
With time gaps (\dtone,\dttwo) = (0, 4 hr), in the $r$-and-$i$ observations case (\emph{right}), 
nearly all transients $r\gtrsim22$ are consistent at a $3\sigma$  level. Furthermore, the distribution is more compact along the $y$-axis, leading to less insightful clues into the physical phenomena. Among these strategies, the $g$-and-$i$ observations with \dtone= 0 and \dttwo= 4 hours time gaps is optimal for recognizing fast transients, allowing us to distinguish most transients  $g~\gtrsim~22$  from SNe Ia and II at $\gtrsim1\sigma$. High resolution PDF versions of this images are available on the project repository \url{https://github.com/fedhere/prestocolor/tree/master/FastTrWP/PhaseSpace/Figures}}}\label{fig:classifier}
\end{center}\end{figure*}

We also include a population of normal Type Ia and Type II/P/L SNe. Together, these represent slower evolving transients from which we wish to \emph{distinguish} our transients of interest. For the SN Ia we include a population of events between $-$7 and $+15$ days of maximum light at a range of redshifts. The SNe Ia SNe are generated with the {\tt SNcosmo} python library \citep{sncosmo}. LSST-like $g$, $r$, and $i$ light curves are generated at redshifts $0.3~\leq~z~\leq~0.7$ using the Spectral Adaptive Lightcurve Template framework {\tt SALT2} \citep[][a light curve fitting model]{salt2}. The three parameters given to {\tt SALT2} are: an observer flux normalization for the spectral templates $x_0$, a light curve shape-luminosity parameter (related to stretch) $x_1$,  and a color parameter $c$ related to the $B - V$ color at peak. Following \citealt{Kessler2010} (figures 2 and 10 therein) we select {\tt SALT2} parameters randomly from Gaussian distributions as $x_1~\sim~N(0,1)$, and set $c~\sim~N(0,0.05)$, and let {\tt SNcosmo} set $x_0$ to produce realistic lightcurves. We generate 100 SNe Ia, constraining the sampling to $-7~\mathrm{days}~<~\mathrm{phase}~<~15~\mathrm{days}$ to peak $B$ brightness. Our rationale to set an upper limit of +15 days is that older transients would have already been discovered by a survey like LSST in previous observations of the same field, so that more data would be available to constraint their characteristics.  Younger SNe, on the other hand, would be ``fast transients.''

For SNe II, we consider a few representative examples: SN~2006bc, SN~2008M SN~2009N, SN~2007od, SN~2008aw~\citep[personal communication]{sncsp}. 


Finally, we consider intermediate luminosity transients (ILOTs) and Calcium rich transients (CARTs) and events from sub-luminous SN Iax. The first two of these models have significantly faster time evolution from type Ia and core collapse SNe, while the Iax subclass exhibit similar properties to SNe Ia using our methodology. These events were simulated for the recently concluded Photometric LSST Astronomical Time-series Classification Challenge~\citep[PLAsTiCC,][]{plasticc}. Light curves were generated assuming the \texttt{minion1016} cadence (see \citealt{COSEP} for details on this strategy)  in the LSST Deep Drilling Fields. We selected these objects over a range of redshifts and Milky Way reddening values, as well as intrinsic parameters of each of these models. 

In order to use these sample light curves to test the effectiveness of various observing strategies we uniformly sample the light curves to a cadence of 30 minutes. Model light curves were generated at this cadence while the light curves from observed transients were fit with low order polynomials and interpolated to a fixed 30 minute cadence within the observed range (no extrapolation). In general, we restrict these template light curves to relatively young objects, interpolating from the \texttt{minion1016} cadence to a 30 minute cadence from $-21$~days to $+15$~days post-maximum.

Each individual transient leads to a family of data-points, corresponding to observations at different epochs in its lifetime.

In \autoref{fig:phasespace}, we plot the location of these sample light curves (excluding the PLAsTiCC transients) in a phase space of observed color (vertical axis) versus observed magnitude change (horizontal axis). In creating this plot, we adopt an observing strategy in which observations are obtained in $g$-and-$i$ band within 30 minutes of each other and a second $g-$band observation is obtained 4 hr later:  (\dtone,\dttwo) = (0.5 hr, 4 hr).
For each individual event the family of data-points generated by observing the transient at different epochs is visible as a `track.' 
Red transients are located at the top of the plot and blue at the bottom. Declining transients are found on the left and rising transients on the right. The infant core-collapse SN found in the declining portion of the plot is the rapidly fading component of the cooling envelope emission observed in SN2016gkg (see \autoref{fig:examples}). With this survey strategy, our transients of interest (colored points) are easily distinguishable from ``normal'' SNe Ia and II (which are clustered near zero magnitude change). At the same time, the observed color provides vital context on the \emph{type} of transient of interest. 

In the section below, we investigate how details of the adopted observing strategy influences our ability to identify transients of interest within this phase space.




\subsection{The sample in the color versus magnitude-change space}\label{sec:classification}

With this data set in hand, we can explore how the color versus magnitude-change space is occupied by normal and unusual transients when two observations are performed in $g$-and-$i$ band or $r$-and-$i$ band at a time interval \dtone\ (in \autoref{fig:phasespace}, \dtone = 30 minutes), and with one further observation in $g-$ or $r-$band \dttwo\ after the first image (in \autoref{fig:phasespace}, \dtone = 4 hr). 

\begin{figure*}
    \begin{center}
    \includegraphics[width=0.9\textwidth]{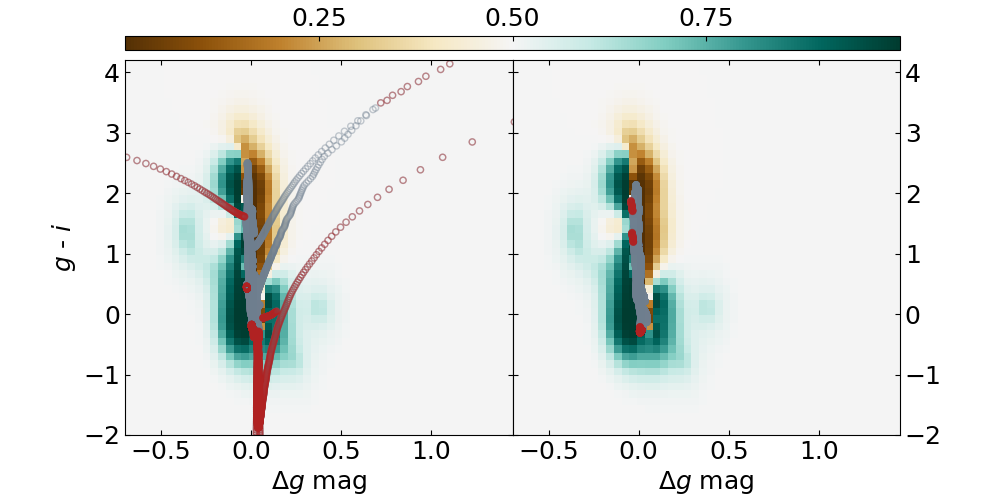}
    \caption{The phase space of color vs. magnitude-change is mapped with the probability of a data point belonging to a fast transient/fast feature, or to a ``normal'' transient. We overlay to this map transients from the PLAsTiCC challenge~\citep{plasticc}: ILOTs and CARTs (\emph{left} panel) and SN Iax (\emph{right} panel). The data points are plotted in red if our simple GP classifier identifies them as ``fast,'' in gray if they are classified as ``normal'' transients.}
    \end{center}
    \label{fig:PLAsTiCC}
\end{figure*}

We train a machine learning classifier to recognize fast transients in this phase space. We use a Gaussian Processes (GP) Probabilistic Classifier \citep{Rasmussen06gaussianprocesses} with a radial basis function (RBF) kernel as implemented in the {\tt scikit-learn} python package~\citep{scikit-learn}, training it to disambiguate fast transients from SNe Ia and II in $g$-and-$i$ and $r$-and-$i$ observations at (\dtone,\dttwo) = (0, 0.5 hr) and (\dtone,\dttwo) = (0.5 hr, 4 hr). The goal of this algorithm is  separating ``normal'' transients from objects that evolve ``fast'': we are approaching this as a binary classification problem (\autoref{fig:classifier} and \autoref{fig:PLAsTiCC}).

We balance these two classes by selecting  $N_\mathrm{fast}$ random observations among the ``normal'' light curves' data points, with $N_\mathrm{fast}$ equal to the number of data points available for all fast transients combined. The ``normal'' sample is split as two-third SN Ia and one-third SN II.  We note that these classes will be significantly unbalanced in real observations, with orders of magnitudes more ``normal'' than unusual objects, but our goal here is to develop a classifier that is \emph{sensitive} or has a high true-positive rate for unusual objects, rather than optimizing for overall accuracy, which would lead to good classification performance on the most common objects, but poor performance on rare events. Additionally, a realistic survey can expect to have GW triggers for KNe, or detections from other wide-field observatories. These further limit the search area on the sky, and constrain the evolution, which further justifies our focus on sensitivity over sample completeness or contamination. Building an optimal photometric classifier that serves the needs of the entire LSST community is a research problem of great interest, and there has been much recent activity on this front \citep[e.g.][]{plasticc}.

The classifier generates a probability for the transient being ``normal'' or ``fast'' anywhere in our phase space.  The goal of using a classifier here is to leverage it as a data-driven way to identify which set of parameters provides the most solid and useful ``not-normal'' label for a newly discovered object. Note that the classifier will default to a 0.5 probability in regions of the space that are not occupied by any of our transients. This does not imply that these regions are uninteresting, and indeed it is in these regions that we might expect to find true outliers: unexpected and unusual, as-of-yet unknown objects!

We train the classifiers with all possible combination of the following time gaps: 
$\dtone~=$~[0.5, ~1.5, ~3.5, ~4, ~4.5, ~6.5]~hr and 
$\dttwo~=$~[0, ~0.5, ~1, ~2]~hr. We do not consider time scales longer than 6.5 hr, due to general consideration on the visibility of a field within a night (see \autoref{sec:implementation}).

The results of our classification exercise are shown in \autoref{fig:classifier}, where red dots are all transients of interest, blue and gold are SNe~Ia and II, gray circles are data points included in the training set, and the classification probability for each point in the phase space is indicated by the background color. The (\dtone,\dttwo) = (0.5 hr, 0.5 hr) strategy classification results are shown in the top panel, and (\dtone,\dttwo) = (0, 4 hr) in the bottom.
On the left-hand side is the result for the filter combination $g$-and-$i$, and on the right for $r$-and-$i$. In all cases we obtain high ($\gtrsim$95\%) accuracy in cross-validation. 

To test the robustness of this classification result, we separately implement a Random Forest Decision Tree Classifier \citep[implemented using scikit-learn,][]{scikit-learn} using the same sample data set of objects as described above on a fiducial set of data using the ($g,i$) filters and (\dtone, \dttwo) = (0.5, 4.5 hr) time gaps which approximate the median case.  Starting with a sparse feature space that only uses the observed color and magnitude difference we find that, as with the GP classifier above, we have a somewhat optimistic classification accuracy of $\gtrsim 97\%$ when using 70\% train and 30\% test data and aggregating across n = 1000 individual decision trees.  

One benefit of tree classification algorithms is that they allow us to straightforwardly interpret the relative importance of our features in determining the assigned class by asking in what fraction of trees is a particular feature at a higher rank in the tree than the others.  When evaluating the relative strengths assigned by the ensemble of trees, we find that the change in magnitude is roughly twice as important as the observed color when assigning a coarse classification ($\sim 0.68$ versus $\sim 0.32$).  

These results lead us to the following considerations:

{\tt { max(\dtone) ---}} The two observations in different filters, separated by \dtone, will be used to assess the color of the transient. 
The fast transients are distinguished mostly due to their $\Delta$mag, but their high rate of change in luminosity implies that only for small \dtone\ we can obtain true color information that would direct our follow-up strategy choices.
Smaller values of \dtone\ --- $\sim$30 minutes --- are better to provide a true estimate of the intrinsic color (see \autoref{sec:classification}). However, values of \dtone\ of up to a few hours, while no longer sensitive to the true ``instantaneous" color of the transient, can still provide some diagnostic power (e.g., of whether a given transient is red or blue), and thus insight into the ongoing physical processes. 


\vspace{1.5mm}
{\tt { min(\dttwo) ---}} The two observations in a matching filter, separated by \dttwo, will be used to assess the rate of brightness change of the transient (the light curve slope). While our classifier successfully separated SNe Ia and II from fast transients in all \dtone-\dttwo\ combinations, with accuracy exceeding 95\%, disambiguation is obviously harder for smaller \dttwo. This is intuitively obvious looking at \autoref{fig:classifier}. 
For \dttwo\ smaller than $\sim4$ hr it is effectively impossible to distinguish fast transients from ``normal'' ones due to the photometric uncertainties, since the classification is mostly driven by the rate of change in luminosity. The uncertainty in $\Delta$mag is plotted for a typical SN Ia (0.3 magnitudes $\mathrm{day}^{-1}$) as observed by LSST with the first measurement at $g/r~=~$20, 22, 24 in the $g$ or $r$ band ($1\sigma$ and $3\sigma$ are shown). At \dttwo\ smaller than $\sim4$ hr nearly all fast transient observations fall within the $1\sigma$ region of type Ia SNe for transients $g~\gtrsim~22$.

\vspace{1.5mm}
{\tt { max(\dttwo)}} --- We advocate for three observations \emph{within the same night} to trigger a follow-up campaign and catch the early, extremely informative stages of a transient’s evolution. In addition, there are general consideration on the visibility of a field within a night to make: $\dttwo\sim4$~hours means we are now observing a field that moved $60\degr$ across the sky: if the first observations happened near zenith the field would now be at a prohibitively high airmass. 

\vspace{1.5mm}
{\bf Filter sequence ---} The repeat observation in the same filter can be obtained either \emph{before or after} the filter pair (while the illustration in \autoref{fig:implementation} always shows it as occurring after). However, given that the main motivation of the {\em Presto-Color} cadence is to identify new objects with rising light curves, we note that beginning the sequence with the image pair is a strategy that is more robust to saturation of very rapidly rising objects: it still provides color and a \emph{slope constraint} even if the final image is saturated. This aspect of the implementation is especially important to consider in light of the relatively faint saturation limit of LSST ($r\sim16$ mag; \citealt{lsst}).
On the other end, beginning the observing sequence with the single visit would lead to a color determination and \emph{lower limits}  to the slope even with an initial non-detection.

\vspace{1.5mm}
{\bf Filter pair $f_1$-$f_2$ ---}
\vspace{-3mm}
\begin{enumerate}
    \item 
The types of energetic, rapidly rising events which the {\em Presto-Color} cadence is designed to find, such as infant SNe, are very blue in color at their earliest phases. This favors the inclusion of LSST filters $g$ or $r$ in the pair (LSST's $u$-band filter's throughput is low, \citealt{oliver2008}).
\item
The magnitude evolution is generally faster at bluer wavelengths, due to rapid cooling which is observed in many of the transients considered here. Thus the transients are better separated along the $\Delta$mag axis when $g-$band observations are repeated.  
\item
Non-adjacent filter pairs provide a larger lever arm to determine the slope of the spectral energy density (SED).
Specifically, the $g$ and $i$ filter choice separates the transients much better than $r$ and $i$ (\autoref{fig:classifier}).  This allows far more insight into the transient from the initial observation triplet.
\end{enumerate}
 
\subsection{Generalization to the PLAsTiCC intermediate luminosity objects}
The simple segregation of the phase space we obtained by training a GP classifier on a few exemplary fast transients/features is not expected to be particularly robust. Nonetheless, we place additional fast transients in this phase space and test the performance of the classifier on these objects, which were not in the training set, to see how generalizable our conclusions are. 

As described in \autoref{sec:transients}, from the PLAsTiCC challenge dataset~\citep{plasticc} we obtained a set of  ILOTs, CARTs, and SNe Iax \citep[][and references therein]{Jha2017}. In this limited sample, ILOTs and CARTs display rapid luminosity evolution, while SNe Iax's behavior is not very different from standard type Ia's. The ILOTs and CARTs are shown in the left panel of \autoref{fig:PLAsTiCC}. The data points are plotted in red if they are classified as ``unusual',' in gray otherwise. ILOTs and CARTs are in part recognized by our classifier, mostly in phases in which their evolution is rapid and the ``tracks'' depart from the middle of the plot along the \emph{x}-axis, but also when their color is blue (bottom of the plot), enforcing the importance of obtaining color information for classification of fast transients. Note that some of the rapid evolution tracks that stray away from the region inhabited by the ``normal'' transients extending in the \emph{red} and \emph{rising} portion of the plot (top right) are not classified as unusual by our classifier simply because we had no points in this region. However, they are in a region of the phase space where the probability is $p~\sim~0.5$, and thus they should be considered interesting. The evolution of Iax SNe, however, is similar to that of normal type Ia's, and they are mostly missed by our classifier. This indicates that our simplistic classification scheme is not sufficient to distinguish SN Iax, or that more observations on longer time scales or contextual data (e.g. proximity to a galaxy center) are required; we remind the reader that photometric classification of transients is a field in rapid evolution \citep[\emph{e.g.}][]{plasticc}.

\section{Conclusions}
We identified minimal requirements that a photometric survey should meet to detect and distinguish unusual, fast-evolving astrophysical transients associated with explosive physics from other transients. We refer to our proposed strategy as {\em Presto-Color} and focus our design on the upcoming LSST. 

The detection and prompt identification of fast transients, such as KNe and rapidly evolving features in type Ia and core-collapse SN, is critical to future synoptic surveys. These transients are poorly observed due to the general inefficiency of synoptic surveys at detecting and recognizing them, and they are poorly understood. Some of these transients are intrinsically rare (KN). Others, while not intrinsically rare, are seldom detected \citep{Drout2014}, as their evolution happens on time-scales shorter than probed by the cadence of ongoing wide-field surveys. They are characterized by rapid luminosity evolution and unusual colors, relative to more common sources. We thus identify the minimal requirements to be three observations in two filters within a night, with a maximum separation for the observations in different filters of $\dtone~\lesssim~ 30$~minutes, which yields an accurate color determination in the presence of a rapidly evolving luminosity, and an optimal repeat of $\dttwo~\gtrsim~ 4$~hr, which provides sufficient leverage on the lightcurve shape to detect the rapid evolution.  

The preferred implementation options for LSST is to use $g$-and-$i$ and repeat the $g$ observation, but alternating between $g$ and $i$ for the repeat filter maximizes observing efficiency (minimizing slew and filter changes). However, we note that LSST is planning to collect nearly twice as many images in $r$ than in $g$\footnote{\url{https://ls.st/srd}} to achieve its four main science goals (\autoref{sec:intro}). If this design feature is preserved in the final LSST observing strategy, {\em Presto-color} could be implemented in  $r$-and-$z$. 
We find that a minimum separation between observations in the same filter of $\gtrsim 1.5$~hr is sufficient to capture a significant change in luminosity only for transients $g\gtrsim22$. 

These two requirements, while not sufficient for accurate studies of individual transients, are sufficient to separate fast transients/features with contemporary machine learning methods. This, in turn, can be used to trigger timely photometric or spectroscopic follow-up, and will lead to valuable rate estimates, as well as observational constraints on the nature of the progenitor systems of these sources. These requirements are only minimal constraints, and still permit significant survey strategy optimization: field selection, sky coverage, and filter selection (provided our constraints) can all be optimized for other science goals.

This research is reproducible. The code is accessible on GitHub\footnote{\url{https://github.com/fedhere/prestocolor}}

\software{scikit-earn \citep{scikit-learn},
        sncosmo \citep{sncosmo},   
          Opsim \citep{opsim},  
          MAF \citep{maf}, 
          astropy \citep{astropy2013, astropy2018}}

\section{Acknowledgments}
This work was developed within the Transients and Variable Stars Science Collaboration (TVS) and the author acknowledges the support of TVS in the preparation of this paper. 

\noindent
The authors acknowledge support from the Flatiron Institute,  Heising-Simons Foundation, and LSST Corporation for the development of this paper.

\noindent
Research support to IA is provided by the GROWTH (Global Relay of Observatories Watching Transients Happen) project funded by the National Science Foundation Partnership in International Research Program under NSF PIRE grant number 1545949.

\noindent 
P.S.C. is grateful for support provided by NASA through the NASA Hubble Fellowship grant \#HST-HF2-51404.001-A awarded by the Space Telescope Science Institute, which is operated by the Association of Universities for Research in Astronomy, Inc., for NASA, under contract NAS 5-26555.

\noindent 
G.N. is supported by the Lasker Fellowship at the Space Telescope Science Institute.

\begin{figure*}[!t]
\begin{center}
\includegraphics[width=6.5cm,height=4.5cm]{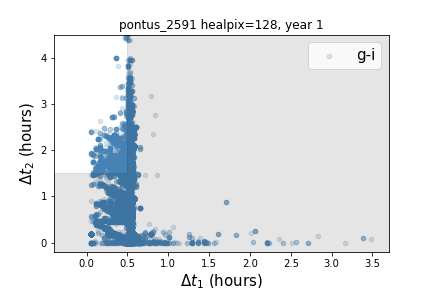}
\includegraphics[width=6.5cm,height=4.5cm]{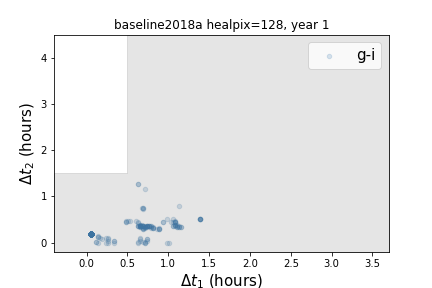}
\includegraphics[width=6.91cm,height=4.5cm]{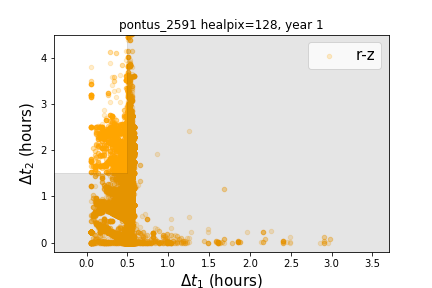}
\includegraphics[width=6.5cm,height=4.5cm]{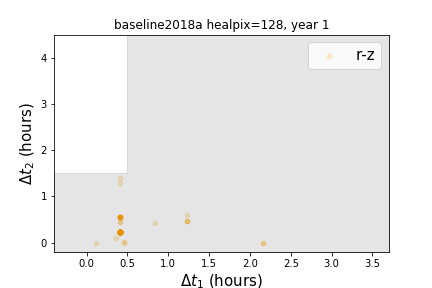}
\includegraphics[width=6.5cm,height=4.5cm]{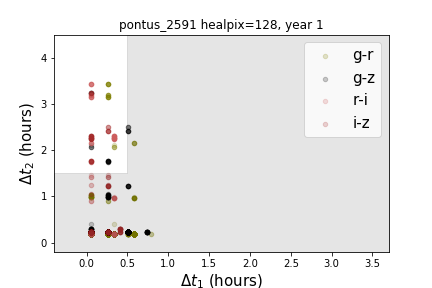}
\includegraphics[width=6.5cm,height=4.5cm]{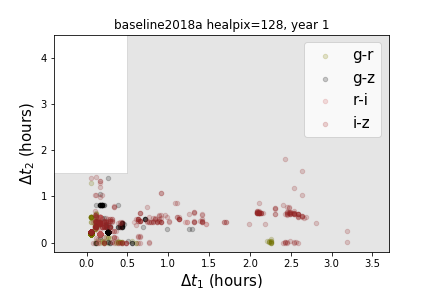}
\caption{{The results of our {\em diagnostic} metric, {\tt threeVisitsWColorMetric}, that checks for fields in {\tt OpSim} LSST strategy simulations that were observed three times in a night with two filters (as labeled) satisfying  constraints on \dtone\ and \dttwo. We explore year-one of the {\tt pontus\_2591} (our {\em Presto-Color} test run; \emph{left} panels), and year-one of the {\tt baseline2018a} LSST {\tt OpSim} runs (\emph{right} panels). We use a {\tt HEALPix} \citep{Gorski05} sky pixelization at a resolution of $nside=128$. All {\tt HEALPix128} ``pixels" that met the conditions of our metric are shown as points, colored by the filter pair $f_1$-and-$f_2$. 
In {\tt pontus\_2591}, $\sim20$ thousand observations in year-one satisfy our constraints strictly ($\dtone<0.5$ and $\dttwo>1.5$) in $g$-and-$i$ and over 40 thousand in $r$-and-$z$.
Very few {\tt HEALPix128} pixels in the {\tt baseline2018a} {\tt OpSim} have observations in triplets in two filters and none that satisfy our constraints.}}\label{fig:metricresult}
\end{center}
\end{figure*}

\appendix 

{\bf LSST Performance Evaluation}

The LSST Project team has created several LSST survey simulations\footnote{\url{http://astro-lsst-01.astro.washington.edu:8080}} in the form of catalogs of observations (including pointing, seeing, filter, etc...) and software that allows users to simulate LSST surveys ({\tt OpSim}, \citealt{opsim}), and to evaluate the performance of LSST simulated survey strategies ({\tt MAF}, \citealt{maf}). Ideally, the success of an LSST strategy simulation in addressing a science case, like the identification of fast transients and fast features in transients, can be conveyed in a single number: a \emph{Figure of Merit} (FoM), so that the overall success of a strategy can be determined by comparing the figures of merit for each science goals. 
The detailed steps involved in evaluating an LSST observing strategy are described in \citet{COSEP}. 
To evaluate a strategy's success in promptly recognizing fast transients/features, we simulated a simple implementation of our proposed WFD strategy and we built two \emph{diagnostic metrics}, described below, to determine the ability of a strategy simulation to collect the observations required for our science. At last, one can extract the relevant observations from the strategy simulation, and analyze them as we would analyze data from LSST. We have not yet performed this last step for this paper.

\subsection{Presto-color OpSim}

The {\tt OpSim} software now runs using a new feature-based algorithm which enables the simulation of a strategy such as the one we have designed. The {\em Presto-Color} survey strategy is accomplished with a combination of three ``Surveys'': the Greedy Algorithm (GA) Survey, the Pairs in different filters (PDF) Survey, and the Pairs in the same filters (PSF) Survey. To combine observations in filters $f_1$ and $f_2$, the Surveys are configured such that when the {\it GA Survey} gets an observation in $f_1$ or $f_2$, the {\it PDF Survey} schedules an observation of the same field with the other filter \dtone = $30 \pm 5$ minutes later, and the {\it PSF Survey} schedules an observation of the same field with the same filter \dttwo = $60 \pm 5$ minutes later (larger \dttwo\ choices should be explored). A test simulation combining $g$-and-$i$ and $r$-and-$z$ was produced to evaluate the impact in performance and test the metrics: this simulation is referred to as {\tt pontus\_2591}. Non-adjacent filters are used to get better leverage on the Spectral Energy Distribution (SED) through color. Overall we noticed that this kind of strategy has a similar throughput (in efficiency and total number of observations) as a strategy to take pairs of observations in different filters.

\subsection{Diagnostic Metrics}

Based on our evaluation of the light curves of fast transients, and the fast features of longer duration transients, we designed two diagnostic metrics and submitted them to the {\em sims\_maf\_contrib}\footnote{\url{https://github.com/LSST-nonproject/sims\_maf\_contrib}} repository.

\begin{enumerate}
\item
{\bf threeVisitsWColorMetric}: a {\it diagnostic} metric that checks if a field was observed three times in a night with two filters, given input constraints on \dtone\ (an upper limit) and \dttwo\ (a lower limit). The specific filter pair of $f_1$ and $f_2$ is also an input to the metric. This metric was run on year-one of {\tt baseline2018a}, the current sample LSST survey strategy, and year-one of our test {\tt OpSim} run {\tt pontus\_2591}. 
\item
{\bf FastTransientMetric}: a {\it diagnostic} metric based on the {\tt Transient Metric} that injects continuous saw-tooth shaped transients with a rising slope (input parameter) and a vertical decline, with peak brightness which can be input independently for each filter (thus enabling the injection of different color transients). The metric calculates the fraction of transients with three detections that are consistent with an input \dtone\ (upper limit) and \dttwo\ (lower limit) and a specific filter-pair.

\end{enumerate}
Results from running {\bf threeVisitsWColorMetric} on the existing LSST {\tt OpSim baseline2018a} simulation and on the new {\tt pontus\_2591} simulation are shown and discussed in \autoref{fig:metricresult}. All LSST observation triplets (all \emph{HEALPix} ---\citealt{Gorski05}---areas of the sky that are observed with triplets) can be plotted on a \dttwo\ versus \dtone\ plane: the $x-$axis is the time between visits in the two different filters (which provides color information), and the $y-$axis is the time between visits in the same filter (which provides lightcurve evolution constraints). The target area for our science is the white region in the plots, where $\dtone~\leq~0.5$ and $\dttwo~\geq~1.5$ hr, such that both color and brightness evolution can be measured, although there is no hard cutoff at 30 minutes in \dtone, so observations just to the left of our target region are valuable, and observations at a larger \dttwo\ within the same night are preferred. While the strategy we designed generates over 20 thousand $g$-and-$i$ and 40 thousand $r$-and-$z$ valuable observations for our science goals, the baseline LSST strategy does not lead to any observations that satisfy our constraints.

\subsection{Figure of Merit}

Our FoM metric is the fraction of events for which the color and rise-time are constrained (within some accuracy). Different science cases will have different input light curves and different gap constraints. We plan on injecting samples of the transients of interest, as well as ``normal" transients, by leveraging the Monte Carlo MAF framework and the {\tt transientLC} metric, recovering the light curve ({\tt PassMetric}) and measuring color and slope, which will be evaluated in the context of  machine learning partitioning of the Color-Slope phase space (\autoref{fig:phasespace} and \autoref{fig:classifier}) to isolate transients for follow-up. Further analysis will assess our ability to distinguish between fast transients with the LSST data alone.


\bibliographystyle{aasjournal}

\begin{thebibliography}{}
\expandafter\ifx\csname natexlab\endcsname\relax\def\natexlab#1{#1}\fi
\providecommand{\url}[1]{\href{#1}{#1}}
\providecommand{\dodoi}[1]{doi:~\href{http://doi.org/#1}{\nolinkurl{#1}}}
\providecommand{\doeprint}[1]{\href{http://ascl.net/#1}{\nolinkurl{http://ascl.net/#1}}}
\providecommand{\doarXiv}[1]{\href{https://arxiv.org/abs/#1}{\nolinkurl{https://arxiv.org/abs/#1}}}

\bibitem[{{Aasi} {et~al.}(2015){Aasi}, {Abadie}, {Abbott}, {Abbott}, {Abbott},
  {Abernathy}, {Accadia}, {Acernese}, {Adams}, {Adams}, \& et~al.}]{Aasi15}
{Aasi}, J., {Abadie}, J., {Abbott}, B.~P., {et~al.} 2015, Classical and Quantum
  Gravity, 32, 115012, \dodoi{10.1088/0264-9381/32/11/115012}

\bibitem[{{Abbott} {et~al.}(2017){Abbott}, {Abbott}, {Abbott}, {Acernese},
  {Ackley}, {Adams}, {Adams}, {Addesso}, {Adhikari}, {Adya}, {Affeldt},
  {Afrough}, {Agarwal}, {Agathos}, {Agatsuma}, {Aggarwal}, {Aguiar}, {Aiello},
  {Ain}, {Ajith}, {Allen}, {Allen}, {Allocca}, {Altin}, {Amato}, {Ananyeva},
  {Anderson}, {Anderson}, {Angelova}, {LIGO Scientific Collaboration}, \&
  {Virgo Collaboration}}]{Abbott2017}
{Abbott}, B.~P., {Abbott}, R., {Abbott}, T.~D., {et~al.} 2017, \prl, 119,
  161101, \dodoi{10.1103/PhysRevLett.119.161101}

\bibitem[{{Acernese} {et~al.}(2015){Acernese}, {Agathos}, {Agatsuma}, {Aisa},
  {Allemandou}, {Allocca}, {Amarni}, {Astone}, {Balestri}, {Ballardin}, \&
  et~al.}]{Acernese15}
{Acernese}, F., {Agathos}, M., {Agatsuma}, K., {et~al.} 2015, Classical and
  Quantum Gravity, 32, 024001, \dodoi{10.1088/0264-9381/32/2/024001}

\bibitem[{{Anderson} {et~al.}(2014){Anderson}, {Gonz{\'a}lez-Gait{\'a}n},
  {Hamuy}, {Guti{\'e}rrez}, {Stritzinger}, {Olivares E.}, {Phillips},
  {Schulze}, {Antezana}, {Bolt}, {Campillay}, {Castell{\'o}n}, {Contreras}, {de
  Jaeger}, {Folatelli}, {F{\"o}rster}, {Freedman}, {Gonz{\'a}lez}, {Hsiao},
  {Krzemi{\'n}ski}, {Krisciunas}, {Maza}, {McCarthy}, {Morrell}, {Persson},
  {Roth}, {Salgado}, {Suntzeff}, \& {Thomas-Osip}}]{sncsp}
{Anderson}, J.~P., {Gonz{\'a}lez-Gait{\'a}n}, S., {Hamuy}, M., {et~al.} 2014,
  \apj, 786, 67, \dodoi{10.1088/0004-637X/786/1/67}

\bibitem[{{Andreoni} {et~al.}(2018){Andreoni}, {Anand}, {Bianco}, {Cenko},
  {Cowperthwaite}, {Coughlin}, {Drout}, {Golkhou}, {Kaplan}, {Mooley},
  {Pritchard}, \& {Singer}}]{Andreoni2018}
{Andreoni}, I., {Anand}, S., {Bianco}, F.~B., {et~al.} 2018, arXiv e-prints,
  arXiv:1812.03161.
\newblock \doarXiv{1812.03161}

\bibitem[{{Arcavi} {et~al.}(2011){Arcavi}, {Gal-Yam}, {Yaron}, {Sternberg},
  {Rabinak}, {Waxman}, {Kasliwal}, {Quimby}, {Ofek}, {Horesh}, {Kulkarni},
  {Filippenko}, {Silverman}, {Cenko}, {Li}, {Bloom}, {Sullivan}, {Nugent},
  {Poznanski}, {Gorbikov}, {Fulton}, {Howell}, {Bersier}, {Riou},
  {Lamotte-Bailey}, {Griga}, {Cohen}, {Hachinger}, {Polishook}, {Xu},
  {Ben-Ami}, {Manulis}, {Walker}, {Maguire}, {Pan}, {Matheson}, {Mazzali},
  {Pian}, {Fox}, {Gehrels}, {Law}, {James}, {Marchant}, {Smith}, {Mottram},
  {Barnsley}, {Kandrashoff}, \& {Clubb}}]{Arcavi2011}
{Arcavi}, I., {Gal-Yam}, A., {Yaron}, O., {et~al.} 2011, \apjl, 742, L18,
  \dodoi{10.1088/2041-8205/742/2/L18}

\bibitem[{{Astropy Collaboration} {et~al.}(2013){Astropy Collaboration},
  {Robitaille}, {Tollerud}, {Greenfield}, {Droettboom}, {Bray}, {Aldcroft},
  {Davis}, {Ginsburg}, {Price-Whelan}, {Kerzendorf}, {Conley}, {Crighton},
  {Barbary}, {Muna}, {Ferguson}, {Grollier}, {Parikh}, {Nair}, {Unther},
  {Deil}, {Woillez}, {Conseil}, {Kramer}, {Turner}, {Singer}, {Fox}, {Weaver},
  {Zabalza}, {Edwards}, {Azalee Bostroem}, {Burke}, {Casey}, {Crawford},
  {Dencheva}, {Ely}, {Jenness}, {Labrie}, {Lim}, {Pierfederici}, {Pontzen},
  {Ptak}, {Refsdal}, {Servillat}, \& {Streicher}}]{astropy2013}
{Astropy Collaboration}, {Robitaille}, T.~P., {Tollerud}, E.~J., {et~al.} 2013,
  \aap, 558, A33, \dodoi{10.1051/0004-6361/201322068}

\bibitem[{Barbary(2014)}]{sncosmo}
Barbary, K. 2014, sncosmo v0.4.2, \dodoi{10.5281/zenodo.11938}.
\newblock \url{https://doi.org/10.5281/zenodo.11938}

\bibitem[{{Bellm}(2014)}]{ztf}
{Bellm}, E. 2014, in The Third Hot-wiring the Transient Universe Workshop, ed.
  P.~R. {Wozniak}, M.~J. {Graham}, A.~A. {Mahabal}, \& R.~{Seaman}, 27--33

\bibitem[{{Berger} {et~al.}(2013){Berger}, {Fong}, \& {Chornock}}]{Berger13}
{Berger}, E., {Fong}, W., \& {Chornock}, R. 2013, \apj, 774, L23,
  \dodoi{10.1088/2041-8205/774/2/L23}

\bibitem[{{Bersten} {et~al.}(2018){Bersten}, {Folatelli}, {Garc{\'{\i}}a}, {van
  Dyk}, {Benvenuto}, {Orellana}, {Buso}, {S{\'a}nchez}, {Tanaka}, {Maeda},
  {Filippenko}, {Zheng}, {Brink}, {Cenko}, {de Jaeger}, {Kumar}, {Moriya},
  {Nomoto}, {Perley}, {Shivvers}, \& {Smith}}]{Bersten2018}
{Bersten}, M.~C., {Folatelli}, G., {Garc{\'{\i}}a}, F., {et~al.} 2018, \nat,
  554, 497, \dodoi{10.1038/nature25151}

\bibitem[{{Bianco} {et~al.}(2011){Bianco}, {Howell}, {Sullivan}, {Conley},
  {Kasen}, {Gonz{\'a}lez-Gait{\'a}n}, {Guy}, {Astier}, {Balland}, {Carlberg},
  {Fouchez}, {Fourmanoit}, {Hardin}, {Hook}, {Lidman}, {Pain},
  {Palanque-Delabrouille}, {Perlmutter}, {Perrett}, {Pritchet}, {Regnault},
  {Rich}, \& {Ruhlmann-Kleider}}]{Bianco11}
{Bianco}, F.~B., {Howell}, D.~A., {Sullivan}, M., {et~al.} 2011, \apj, 741, 20,
  \dodoi{10.1088/0004-637X/741/1/20}

\bibitem[{{Blagorodnova} {et~al.}(2017){Blagorodnova}, {Kotak}, {Polshaw},
  {Kasliwal}, {Cao}, {Cody}, {Doran}, {Elias-Rosa}, {Fraser}, {Fremling},
  {Gonzalez-Fernandez}, {Harmanen}, {Jencson}, {Kankare}, {Kudritzki},
  {Kulkarni}, {Magnier}, {Manulis}, {Masci}, {Mattila}, {Nugent}, {Ochner},
  {Pastorello}, {Reynolds}, {Smith}, {Sollerman}, {Taddia}, {Terreran},
  {Tomasella}, {Turatto}, {Vreeswijk}, {Wozniak}, \&
  {Zaggia}}]{Blagorodnova2017}
{Blagorodnova}, N., {Kotak}, R., {Polshaw}, J., {et~al.} 2017, \apj, 834, 107,
  \dodoi{10.3847/1538-4357/834/2/107}

\bibitem[{{Bricman} \& {Gomboc}(2018)}]{Bricman2018}
{Bricman}, K., \& {Gomboc}, A. 2018, arXiv e-prints, arXiv:1812.06054.
\newblock \doarXiv{1812.06054}

\bibitem[{{Cowperthwaite} {et~al.}(2017){Cowperthwaite}, {Berger}, {Villar},
  {Metzger}, {Nicholl}, {Chornock}, {Blanchard}, {Fong}, {Margutti},
  {Soares-Santos}, {Alexander}, {Allam}, {Annis}, {Brout}, {Brown}, {Butler},
  {Chen}, {Diehl}, {Doctor}, {Drout}, {Eftekhari}, {Farr}, {Finley}, {Foley},
  {Frieman}, {Fryer}, {Garc{\'\i}a-Bellido}, {Gill}, {Guillochon}, {Herner},
  {Holz}, {Kasen}, {Kessler}, {Marriner}, {Matheson}, {Neilsen}, {Quataert},
  {Palmese}, {Rest}, {Sako}, {Scolnic}, {Smith}, {Tucker}, {Williams},
  {Balbinot}, {Carlin}, {Cook}, {Durret}, {Li}, {Lopes}, {Louren{\c{c}}o},
  {Marshall}, {Medina}, {Muir}, {Mu{\~n}oz}, {Sauseda}, {Schlegel}, {Secco},
  {Vivas}, {Wester}, {Zenteno}, {Zhang}, {Abbott}, {Banerji}, {Bechtol},
  {Benoit-L{\'e}vy}, {Bertin}, {Buckley-Geer}, {Burke}, {Capozzi}, {Carnero
  Rosell}, {Carrasco Kind}, {Castander}, {Crocce}, {Cunha}, {D'Andrea}, {da
  Costa}, {Davis}, {DePoy}, {Desai}, {Dietrich}, {Drlica-Wagner}, {Eifler},
  {Evrard}, {Fernandez}, {Flaugher}, {Fosalba}, {Gaztanaga}, {Gerdes},
  {Giannantonio}, {Goldstein}, {Gruen}, {Gruendl}, {Gutierrez}, {Honscheid},
  {Jain}, {James}, {Jeltema}, {Johnson}, {Johnson}, {Kent}, {Krause}, {Kron},
  {Kuehn}, {Nuropatkin}, {Lahav}, {Lima}, {Lin}, {Maia}, {March}, {Martini},
  {McMahon}, {Menanteau}, {Miller}, {Miquel}, {Mohr}, {Neilsen}, {Nichol},
  {Ogando}, {Plazas}, {Roe}, {Romer}, {Roodman}, {Rykoff}, {Sanchez},
  {Scarpine}, {Schindler}, {Schubnell}, {Sevilla-Noarbe}, {Smith}, {Smith},
  {Sobreira}, {Suchyta}, {Swanson}, {Tarle}, {Thomas}, {Thomas}, {Troxel},
  {Vikram}, {Walker}, {Wechsler}, {Weller}, {Yanny}, \& {Zuntz}}]{Cowp+17}
{Cowperthwaite}, P.~S., {Berger}, E., {Villar}, V.~A., {et~al.} 2017, \apj,
  848, L17, \dodoi{10.3847/2041-8213/aa8fc7}

\bibitem[{{D'Andrea}(2011)}]{2011PhDT........37D}
{D'Andrea}, C.~B. 2011, PhD thesis, University of Pennsylvania

\bibitem[{{De} {et~al.}(2018){De}, {Kasliwal}, {Ofek}, {Moriya}, {Burke},
  {Cao}, {Cenko}, {Doran}, {Duggan}, {Fender}, {Fransson}, {Gal-Yam}, {Horesh},
  {Kulkarni}, {Laher}, {Lunnan}, {Manulis}, {Masci}, {Mazzali}, {Nugent},
  {Perley}, {Petrushevska}, {Piro}, {Rumsey}, {Sollerman}, {Sullivan}, \&
  {Taddia}}]{De2018}
{De}, K., {Kasliwal}, M.~M., {Ofek}, E.~O., {et~al.} 2018, Science, 362, 201,
  \dodoi{10.1126/science.aas8693}

\bibitem[{{Delgado} {et~al.}(2014){Delgado}, {Saha}, {Chandrasekharan}, {Cook},
  {Petry}, \& {Ridgway}}]{opsim}
{Delgado}, F., {Saha}, A., {Chandrasekharan}, S., {et~al.} 2014, in \procspie,
  Vol. 9150, Modeling, Systems Engineering, and Project Management for
  Astronomy VI, 915015

\bibitem[{{Drout} {et~al.}(2013){Drout}, {Soderberg}, {Mazzali}, {Parrent},
  {Margutti}, {Milisavljevic}, {Sanders}, {Chornock}, {Foley}, {Kirshner},
  {Filippenko}, {Li}, {Brown}, {Cenko}, {Chakraborti}, {Challis}, {Friedman},
  {Ganeshalingam}, {Hicken}, {Jensen}, {Modjaz}, {Perets}, {Silverman}, \&
  {Wong}}]{Drout2013}
{Drout}, M.~R., {Soderberg}, A.~M., {Mazzali}, P.~A., {et~al.} 2013, \apj, 774,
  58, \dodoi{10.1088/0004-637X/774/1/58}

\bibitem[{{Drout} {et~al.}(2014){Drout}, {Chornock}, {Soderberg}, {Sanders},
  {McKinnon}, {Rest}, {Foley}, {Milisavljevic}, {Margutti}, {Berger},
  {Calkins}, {Fong}, {Gezari}, {Huber}, {Kankare}, {Kirshner}, {Leibler},
  {Lunnan}, {Mattila}, {Marion}, {Narayan}, {Riess}, {Roth}, {Scolnic},
  {Smartt}, {Tonry}, {Burgett}, {Chambers}, {Hodapp}, {Jedicke}, {Kaiser},
  {Magnier}, {Metcalfe}, {Morgan}, {Price}, \& {Waters}}]{Drout2014}
{Drout}, M.~R., {Chornock}, R., {Soderberg}, A.~M., {et~al.} 2014, \apj, 794,
  23, \dodoi{10.1088/0004-637X/794/1/23}

\bibitem[{{Drout} {et~al.}(2017){Drout}, {Piro}, {Shappee}, {Kilpatrick},
  {Simon}, {Contreras}, {Coulter}, {Foley}, {Siebert}, {Morrell}, {Boutsia},
  {Di Mille}, {Holoien}, {Kasen}, {Kollmeier}, {Madore}, {Monson},
  {Murguia-Berthier}, {Pan}, {Prochaska}, {Ramirez-Ruiz}, {Rest}, {Adams},
  {Alatalo}, {Ba{\~n}ados}, {Baughman}, {Beers}, {Bernstein}, {Bitsakis},
  {Campillay}, {Hansen}, {Higgs}, {Ji}, {Maravelias}, {Marshall}, {Moni Bidin},
  {Prieto}, {Rasmussen}, {Rojas-Bravo}, {Strom}, {Ulloa},
  {Vargas-Gonz{\'a}lez}, {Wan}, \& {Whitten}}]{Drout2017}
{Drout}, M.~R., {Piro}, A.~L., {Shappee}, B.~J., {et~al.} 2017, Science, 358,
  1570, \dodoi{10.1126/science.aaq0049}

\bibitem[{{Fuller}(2017)}]{Fuller2017}
{Fuller}, J. 2017, \mnras, 470, 1642, \dodoi{10.1093/mnras/stx1314}

\bibitem[{{Gezari} {et~al.}(2018){Gezari}, {van Velzen}, {Hung}, {Cenko}, \&
  {Arcavi}}]{Gezari2018}
{Gezari}, S., {van Velzen}, S., {Hung}, T., {Cenko}, B., \& {Arcavi}, I. 2018,
  arXiv e-prints, arXiv:1812.07036.
\newblock \doarXiv{1812.07036}

\bibitem[{{G{\'o}rski} {et~al.}(2005){G{\'o}rski}, {Hivon}, {Banday},
  {Wandelt}, {Hansen}, {Reinecke}, \& {Bartelmann}}]{Gorski05}
{G{\'o}rski}, K.~M., {Hivon}, E., {Banday}, A.~J., {et~al.} 2005, \apj, 622,
  759, \dodoi{10.1086/427976}

\bibitem[{{Gr{\"a}fener} {et~al.}(2012){Gr{\"a}fener}, {Owocki}, \&
  {Vink}}]{Grafener2012}
{Gr{\"a}fener}, G., {Owocki}, S.~P., \& {Vink}, J.~S. 2012, Astronomy and
  Astrophysics, 538, A40, \dodoi{10.1051/0004-6361/201117497}

\bibitem[{{Graham} {et~al.}(2015){Graham}, {Foley}, {Zheng}, {Kelly},
  {Shivvers}, {Silverman}, {Filippenko}, {Clubb}, \&
  {Ganeshalingam}}]{Graham2015}
{Graham}, M.~L., {Foley}, R.~J., {Zheng}, W., {et~al.} 2015, \mnras, 446, 2073,
  \dodoi{10.1093/mnras/stu2221}

\bibitem[{{Guy} {et~al.}(2007){Guy}, {Astier}, {Baumont}, {Hardin}, {Pain},
  {Regnault}, {Basa}, {Carlberg}, {Conley}, {Fabbro}, {Fouchez}, {Hook},
  {Howell}, {Perrett}, {Pritchet}, {Rich}, {Sullivan}, {Antilogus}, {Aubourg},
  {Bazin}, {Bronder}, {Filiol}, {Palanque-Delabrouille}, {Ripoche}, \&
  {Ruhlmann-Kleider}}]{salt2}
{Guy}, J., {Astier}, P., {Baumont}, S., {et~al.} 2007, \aap, 466, 11,
  \dodoi{10.1051/0004-6361:20066930}

\bibitem[{{Hayden}(2013)}]{2013PhDT.......326H}
{Hayden}, B.~T. 2013, PhD thesis, University of Notre Dame

\bibitem[{{Hayden} {et~al.}(2010){Hayden}, {Garnavich}, {Kasen}, {Dilday},
  {Frieman}, {Jha}, {Lampeitl}, {Nichol}, {Sako}, {Schneider}, {Smith},
  {Sollerman}, \& {Wheeler}}]{Hayden2010}
{Hayden}, B.~T., {Garnavich}, P.~M., {Kasen}, D., {et~al.} 2010, \apj, 722,
  1691, \dodoi{10.1088/0004-637X/722/2/1691}

\bibitem[{{Hillebrandt} \& {Niemeyer}(2000)}]{Hillebrandt2000}
{Hillebrandt}, W., \& {Niemeyer}, J.~C. 2000, Annual Review of Astronomy and
  Astrophysics, 38, 191, \dodoi{10.1146/annurev.astro.38.1.191}

\bibitem[{{Hosseinzadeh} {et~al.}(2017){Hosseinzadeh}, {Sand}, {Valenti},
  {Brown}, {Howell}, {McCully}, {Kasen}, {Arcavi}, {Azalee Bostroem},
  {Tartaglia}, {Hsiao}, {Davis}, {Shahbandeh}, \&
  {Stritzinger}}]{Hosseinzadeh2017}
{Hosseinzadeh}, G., {Sand}, D.~J., {Valenti}, S., {et~al.} 2017, \apjl, 845,
  L11, \dodoi{10.3847/2041-8213/aa8402}

\bibitem[{{Ivezi{\'c}} {et~al.}(2008){Ivezi{\'c}}, {Kahn}, {Tyson}, {Abel},
  {Acosta}, {Allsman}, {Alonso}, {AlSayyad}, {Anderson}, {Andrew}, \&
  et~al.}]{lsst}
{Ivezi{\'c}}, {\v Z}., {Kahn}, S.~M., {Tyson}, J.~A., {et~al.} 2008, ArXiv
  e-prints.
\newblock \doarXiv{0805.2366}

\bibitem[{{Jha}(2017)}]{Jha2017}
{Jha}, S.~W. 2017, {Type Iax Supernovae} (Springer International Publishing
  AG), 375

\bibitem[{{Jin} {et~al.}(2015){Jin}, {Li}, {Cano}, {Covino}, {Fan}, \&
  {Wei}}]{Jin15}
{Jin}, Z.-P., {Li}, X., {Cano}, Z., {et~al.} 2015, \apj, 811, L22,
  \dodoi{10.1088/2041-8205/811/2/L22}

\bibitem[{{Jones} {et~al.}(2014){Jones}, {Yoachim}, {Chandrasekharan},
  {Connolly}, {Cook}, {Ivezic}, {Krughoff}, {Petry}, \& {Ridgway}}]{maf}
{Jones}, R.~L., {Yoachim}, P., {Chandrasekharan}, S., {et~al.} 2014, in
  \procspie, Vol. 9149, Observatory Operations: Strategies, Processes, and
  Systems V, 91490B

\bibitem[{{Kasen}(2010)}]{Kasen2010}
{Kasen}, D. 2010, \apj, 708, 1025, \dodoi{10.1088/0004-637X/708/2/1025}

\bibitem[{{Kasen} {et~al.}(2015){Kasen}, {Fern{\'a}ndez}, \&
  {Metzger}}]{Kasen2015}
{Kasen}, D., {Fern{\'a}ndez}, R., \& {Metzger}, B.~D. 2015, \mnras, 450, 1777,
  \dodoi{10.1093/mnras/stv721}

\bibitem[{{Kashiyama} \& {Quataert}(2015)}]{Kashiyama2015}
{Kashiyama}, K., \& {Quataert}, E. 2015, \mnras, 451, 2656,
  \dodoi{10.1093/mnras/stv1164}

\bibitem[{{Kasliwal} {et~al.}(2010){Kasliwal}, {Kulkarni}, {Gal-Yam}, {Yaron},
  {Quimby}, {Ofek}, {Nugent}, {Poznanski}, {Jacobsen}, {Sternberg}, {Arcavi},
  {Howell}, {Sullivan}, {Rich}, {Burke}, {Brimacombe}, {Milisavljevic},
  {Fesen}, {Bildsten}, {Shen}, {Cenko}, {Bloom}, {Hsiao}, {Law}, {Gehrels},
  {Immler}, {Dekany}, {Rahmer}, {Hale}, {Smith}, {Zolkower}, {Velur},
  {Walters}, {Henning}, {Bui}, \& {McKenna}}]{Kasliwal2010}
{Kasliwal}, M.~M., {Kulkarni}, S.~R., {Gal-Yam}, A., {et~al.} 2010, \apj, 723,
  L98, \dodoi{10.1088/2041-8205/723/1/L98}

\bibitem[{{Kasliwal} {et~al.}(2017){Kasliwal}, {Nakar}, {Singer}, {Kaplan},
  {Cook}, {Van Sistine}, {Lau}, {Fremling}, {Gottlieb}, {Jencson}, {Adams},
  {Feindt}, {Hotokezaka}, {Ghosh}, {Perley}, {Yu}, {Piran}, {Allison},
  {Anupama}, {Balasubramanian}, {Bannister}, {Bally}, {Barnes}, {Barway},
  {Bellm}, {Bhalerao}, {Bhattacharya}, {Blagorodnova}, {Bloom}, {Brady},
  {Cannella}, {Chatterjee}, {Cenko}, {Cobb}, {Copperwheat}, {Corsi}, {De},
  {Dobie}, {Emery}, {Evans}, {Fox}, {Frail}, {Frohmaier}, {Goobar}, {Hallinan},
  {Harrison}, {Helou}, {Hinderer}, {Ho}, {Horesh}, {Ip}, {Itoh}, {Kasen},
  {Kim}, {Kuin}, {Kupfer}, {Lynch}, {Madsen}, {Mazzali}, {Miller}, {Mooley},
  {Murphy}, {Ngeow}, {Nichols}, {Nissanke}, {Nugent}, {Ofek}, {Qi}, {Quimby},
  {Rosswog}, {Rusu}, {Sadler}, {Schmidt}, {Sollerman}, {Steele}, {Williamson},
  {Xu}, {Yan}, {Yatsu}, {Zhang}, \& {Zhao}}]{Kasliwal2017}
{Kasliwal}, M.~M., {Nakar}, E., {Singer}, L.~P., {et~al.} 2017, Science, 358,
  1559, \dodoi{10.1126/science.aap9455}

\bibitem[{Kessler {et~al.}(2010)Kessler, Cinabro, Bassett, Dilday, Frieman,
  Garnavich, Jha, Marriner, Nichol, Sako, Smith, Bernstein, Bizyaev, Goobar,
  Kuhlmann, Schneider, \& Stritzinger}]{Kessler2010}
Kessler, R., Cinabro, D., Bassett, B., {et~al.} 2010, The Astrophysical
  Journal, 717, 40, \dodoi{10.1088/0004-637x/717/1/40}

\bibitem[{{Khazov} {et~al.}(2016){Khazov}, {Yaron}, {Gal-Yam}, {Manulis},
  {Rubin}, {Kulkarni}, {Arcavi}, {Kasliwal}, {Ofek}, {Cao}, {Perley},
  {Sollerman}, {Horesh}, {Sullivan}, {Filippenko}, {Nugent}, {Howell}, {Cenko},
  {Silverman}, {Ebeling}, {Taddia}, {Johansson}, {Laher}, {Surace},
  {Rebbapragada}, {Wozniak}, \& {Matheson}}]{Khazov2016}
{Khazov}, D., {Yaron}, O., {Gal-Yam}, A., {et~al.} 2016, \apj, 818, 3,
  \dodoi{10.3847/0004-637X/818/1/3}

\bibitem[{{Li} \& {Paczy{\'n}ski}(1998)}]{Li1998}
{Li}, L.-X., \& {Paczy{\'n}ski}, B. 1998, \apj, 507, L59,
  \dodoi{10.1086/311680}

\bibitem[{{LSST Science Collaboration} {et~al.}(2017){LSST Science
  Collaboration}, {Marshall}, {Anguita}, {Bianco}, {Bellm}, {Brandt},
  {Clarkson}, {Connolly}, {Gawiser}, {Ivezic}, {Jones}, {Lochner}, {Lund},
  {Mahabal}, {Nidever}, {Olsen}, {Ridgway}, {Rhodes}, {Shemmer}, {Trilling},
  {Vivas}, {Walkowicz}, {Willman}, {Yoachim}, {Anderson}, {Antilogus}, {Angus},
  {Arcavi}, {Awan}, {Biswas}, {Bell}, {Bennett}, {Britt}, {Buzasi},
  {Casetti-Dinescu}, {Chomiuk}, {Claver}, {Cook}, {Davenport}, {Debattista},
  {Digel}, {Doctor}, {Firth}, {Foley}, {Fong}, {Galbany}, {Giampapa}, {Gizis},
  {Graham}, {Grillmair}, {Gris}, {Haiman}, {Hartigan}, {Hawley}, {Hlozek},
  {Jha}, {Johns-Krull}, {Kanbur}, {Kalogera}, {Kashyap}, {Kasliwal}, {Kessler},
  {Kim}, {Kurczynski}, {Lahav}, {Liu}, {Malz}, {Margutti}, {Matheson},
  {McEwen}, {McGehee}, {Meibom}, {Meyers}, {Monet}, {Neilsen}, {Newman},
  {O'Dowd}, {Peiris}, {Penny}, {Peters}, {Poleski}, {Ponder}, {Richards},
  {Rho}, {Rubin}, {Schmidt}, {Schuhmann}, {Shporer}, {Slater}, {Smith},
  {Soares-Santos}, {Stassun}, {Strader}, {Strauss}, {Street}, {Stubbs},
  {Sullivan}, {Szkody}, {Trimble}, {Tyson}, {de Val-Borro}, {Valenti},
  {Wagoner}, {Wood-Vasey}, \& {Zauderer}}]{COSEP}
{LSST Science Collaboration}, {Marshall}, P., {Anguita}, T., {et~al.} 2017,
  ArXiv e-prints.
\newblock \doarXiv{1708.04058}

\bibitem[{{Margutti} {et~al.}(2018){Margutti}, {Metzger}, {Chornock}, {Vurm},
  {Roth}, {Grefenstette}, {Savchenko}, {Cartier}, {Steiner}, {Terreran},
  {Migliori}, {Milisavljevic}, {Alexander}, {Bietenholz}, {Blanchard}, {Bozzo},
  {Brethauer}, {Chilingarian}, {Coppejans}, {Ducci}, {Ferrigno}, {Fong},
  {G{\"O}tz}, {Guidorzi}, {Hajela}, {Hurley}, {Kuulkers}, {Laurent},
  {Mereghetti}, {Nicholl}, {Patnaude}, {Ubertini}, {Banovetz}, {Bartel},
  {Berger}, {Coughlin}, {Eftekhari}, {Frederiks}, {Kozlova}, {Laskar},
  {Svinkin}, {Drout}, {Macfadyen}, \& {Paterson}}]{Margutti2018}
{Margutti}, R., {Metzger}, B.~D., {Chornock}, R., {et~al.} 2018, ArXiv
  e-prints, arXiv:1810.10720.
\newblock \doarXiv{1810.10720}

\bibitem[{{Matzner} {et~al.}(2013){Matzner}, {Levin}, \& {Ro}}]{Matzner2013}
{Matzner}, C.~D., {Levin}, Y., \& {Ro}, S. 2013, \apj, 779, 60,
  \dodoi{10.1088/0004-637X/779/1/60}

\bibitem[{{Metzger}(2017)}]{MetzgerKN}
{Metzger}, B.~D. 2017, Living Reviews in Relativity, 20, 3,
  \dodoi{10.1007/s41114-017-0006-z}

\bibitem[{{Metzger} {et~al.}(2009){Metzger}, {Piro}, {Quataert}, \&
  {Thompson}}]{Metzger2009}
{Metzger}, B.~D., {Piro}, A.~L., {Quataert}, E., \& {Thompson}, T.~A. 2009,
  ArXiv e-prints, arXiv:0908.1127.
\newblock \doarXiv{0908.1127}

\bibitem[{{Metzger} {et~al.}(2018){Metzger}, {Thompson}, \&
  {Quataert}}]{Metzger2018}
{Metzger}, B.~D., {Thompson}, T.~A., \& {Quataert}, E. 2018, \apj, 856, 101,
  \dodoi{10.3847/1538-4357/aab095}

\bibitem[{{Metzger} {et~al.}(2010){Metzger}, {Mart{\'\i}nez-Pinedo}, {Darbha},
  {Quataert}, {Arcones}, {Kasen}, {Thomas}, {Nugent}, {Panov}, \&
  {Zinner}}]{Metzger2010}
{Metzger}, B.~D., {Mart{\'\i}nez-Pinedo}, G., {Darbha}, S., {et~al.} 2010,
  \mnras, 406, 2650, \dodoi{10.1111/j.1365-2966.2010.16864.x}

\bibitem[{{Modjaz} {et~al.}(2009){Modjaz}, {Li}, {Butler}, {Chornock},
  {Perley}, {Blondin}, {Bloom}, {Filippenko}, {Kirshner}, {Kocevski},
  {Poznanski}, {Hicken}, {Foley}, {Stringfellow}, {Berlind}, {Barrado y
  Navascues}, {Blake}, {Bouy}, {Brown}, {Challis}, {Chen}, {de Vries},
  {Dufour}, {Falco}, {Friedman}, {Ganeshalingam}, {Garnavich}, {Holden},
  {Illingworth}, {Lee}, {Liebert}, {Marion}, {Olivier}, {Prochaska},
  {Silverman}, {Smith}, {Starr}, {Steele}, {Stockton}, {Williams}, \&
  {Wood-Vasey}}]{Modjaz2009}
{Modjaz}, M., {Li}, W., {Butler}, N., {et~al.} 2009, \apj, 702, 226,
  \dodoi{10.1088/0004-637X/702/1/226}

\bibitem[{{Moriya} {et~al.}(2010){Moriya}, {Tominaga}, {Tanaka}, {Nomoto},
  {Sauer}, {Mazzali}, {Maeda}, \& {Suzuki}}]{Moriya2010}
{Moriya}, T., {Tominaga}, N., {Tanaka}, M., {et~al.} 2010, \apj, 719, 1445,
  \dodoi{10.1088/0004-637X/719/2/1445}

\bibitem[{{Nakar} \& {Piro}(2014)}]{Nakar2014}
{Nakar}, E., \& {Piro}, A.~L. 2014, \apj, 788, 193,
  \dodoi{10.1088/0004-637X/788/2/193}

\bibitem[{{Nakar} \& {Sari}(2010)}]{Nakar2010}
{Nakar}, E., \& {Sari}, R. 2010, \apj, 725, 904,
  \dodoi{10.1088/0004-637X/725/1/904}

\bibitem[{{Nugent} {et~al.}(2011){Nugent}, {Sullivan}, {Cenko}, {Thomas},
  {Kasen}, {Howell}, {Bersier}, {Bloom}, {Kulkarni}, {Kandrashoff},
  {Filippenko}, {Silverman}, {Marcy}, {Howard}, {Isaacson}, {Maguire},
  {Suzuki}, {Tarlton}, {Pan}, {Bildsten}, {Fulton}, {Parrent}, {Sand},
  {Podsiadlowski}, {Bianco}, {Dilday}, {Graham}, {Lyman}, {James}, {Kasliwal},
  {Law}, {Quimby}, {Hook}, {Walker}, {Mazzali}, {Pian}, {Ofek}, {Gal-Yam}, \&
  {Poznanski}}]{Nugent2011}
{Nugent}, P.~E., {Sullivan}, M., {Cenko}, S.~B., {et~al.} 2011, \nat, 480, 344,
  \dodoi{10.1038/nature10644}

\bibitem[{{Olivier} {et~al.}(2008){Olivier}, {Seppala}, \&
  {Gilmore}}]{oliver2008}
{Olivier}, S.~S., {Seppala}, L., \& {Gilmore}, K. 2008, in \procspie, Vol.
  7018, Advanced Optical and Mechanical Technologies in Telescopes and
  Instrumentation, 70182G

\bibitem[{Pedregosa {et~al.}(2011)Pedregosa, Varoquaux, Gramfort, Michel,
  Thirion, Grisel, Blondel, Prettenhofer, Weiss, Dubourg, Vanderplas, Passos,
  Cournapeau, Brucher, Perrot, \& Duchesnay}]{scikit-learn}
Pedregosa, F., Varoquaux, G., Gramfort, A., {et~al.} 2011, Journal of Machine
  Learning Research, 12, 2825

\bibitem[{{Piro} \& {Kollmeier}(2018)}]{Piro2018}
{Piro}, A.~L., \& {Kollmeier}, J.~A. 2018, \apj, 855, 103,
  \dodoi{10.3847/1538-4357/aaaab3}

\bibitem[{{Piro} \& {Morozova}(2016)}]{Piro2016}
{Piro}, A.~L., \& {Morozova}, V.~S. 2016, \apj, 826, 96,
  \dodoi{10.3847/0004-637X/826/1/96}

\bibitem[{{Polin} {et~al.}(2018){Polin}, {Nugent}, \& {Kasen}}]{Polin2018}
{Polin}, A., {Nugent}, P., \& {Kasen}, D. 2018, ArXiv e-prints,
  arXiv:1811.07127.
\newblock \doarXiv{1811.07127}

\bibitem[{{Prentice} {et~al.}(2018){Prentice}, {Maguire}, {Smartt}, {Magee},
  {Schady}, {Sim}, {Chen}, {Clark}, {Colin}, {Fulton}, {McBrien},
  {O{\textquoteright}Neill}, {Smith}, {Ashall}, {Chambers}, {Denneau},
  {Flewelling}, {Heinze}, {Holoien}, {Huber}, {Kochanek}, {Mazzali}, {Prieto},
  {Rest}, {Shappee}, {Stalder}, {Stanek}, {Stritzinger}, {Thompson}, \&
  {Tonry}}]{Prentice2018}
{Prentice}, S.~J., {Maguire}, K., {Smartt}, S.~J., {et~al.} 2018, \apj, 865,
  L3, \dodoi{10.3847/2041-8213/aadd90}

\bibitem[{{Price-Whelan} {et~al.}(2018){Price-Whelan}, {Sip{\H{o}}cz},
  {G{\"u}nther}, {Lim}, {Crawford}, {Conseil}, {Shupe}, {Craig}, {Dencheva},
  {Ginsburg}, {VanderPlas}, {Bradley}, {P{\'e}rez-Su{\'a}rez}, {de Val-Borro},
  {Paper Contributors}, {Aldcroft}, {Cruz}, {Robitaille}, {Tollerud},
  {Coordination Committee}, {Ardelean}, {Babej}, {Bach}, {Bachetti}, {Bakanov},
  {Bamford}, {Barentsen}, {Barmby}, {Baumbach}, {Berry}, {Biscani}, {Boquien},
  {Bostroem}, {Bouma}, {Brammer}, {Bray}, {Breytenbach}, {Buddelmeijer},
  {Burke}, {Calderone}, {Cano Rodr{\'\i}guez}, {Cara}, {Cardoso}, {Cheedella},
  {Copin}, {Corrales}, {Crichton}, {D{\textquoteright}Avella}, {Deil},
  {Depagne}, {Dietrich}, {Donath}, {Droettboom}, {Earl}, {Erben}, {Fabbro},
  {Ferreira}, {Finethy}, {Fox}, {Garrison}, {Gibbons}, {Goldstein}, {Gommers},
  {Greco}, {Greenfield}, {Groener}, {Grollier}, {Hagen}, {Hirst}, {Homeier},
  {Horton}, {Hosseinzadeh}, {Hu}, {Hunkeler}, {Ivezi{\'c}}, {Jain}, {Jenness},
  {Kanarek}, {Kendrew}, {Kern}, {Kerzendorf}, {Khvalko}, {King}, {Kirkby},
  {Kulkarni}, {Kumar}, {Lee}, {Lenz}, {Littlefair}, {Ma}, {Macleod},
  {Mastropietro}, {McCully}, {Montagnac}, {Morris}, {Mueller}, {Mumford},
  {Muna}, {Murphy}, {Nelson}, {Nguyen}, {Ninan}, {N{\"o}the}, {Ogaz}, {Oh},
  {Parejko}, {Parley}, {Pascual}, {Patil}, {Patil}, {Plunkett}, {Prochaska},
  {Rastogi}, {Reddy Janga}, {Sabater}, {Sakurikar}, {Seifert}, {Sherbert},
  {Sherwood-Taylor}, {Shih}, {Sick}, {Silbiger}, {Singanamalla}, {Singer},
  {Sladen}, {Sooley}, {Sornarajah}, {Streicher}, {Teuben}, {Thomas},
  {Tremblay}, {Turner}, {Terr{\'o}n}, {van Kerkwijk}, {de la Vega}, {Watkins},
  {Weaver}, {Whitmore}, {Woillez}, {Zabalza}, \& {Contributors}}]{astropy2018}
{Price-Whelan}, A.~M., {Sip{\H{o}}cz}, B.~M., {G{\"u}nther}, H.~M., {et~al.}
  2018, \aj, 156, 123, \dodoi{10.3847/1538-3881/aabc4f}

\bibitem[{{Pursiainen} {et~al.}(2018){Pursiainen}, {Childress}, {Smith},
  {Prajs}, {Sullivan}, {Davis}, {Foley}, {Asorey}, {Calcino}, {Carollo},
  {Curtin}, {D'Andrea}, {Glazebrook}, {Gutierrez}, {Hinton}, {Hoormann},
  {Inserra}, {Kessler}, {King}, {Kuehn}, {Lewis}, {Lidman}, {Macaulay},
  {M{\"o}ller}, {Nichol}, {Sako}, {Sommer}, {Swann}, {Tucker}, {Uddin},
  {Wiseman}, {Zhang}, {Abbott}, {Abdalla}, {Allam}, {Annis}, {Avila}, {Brooks},
  {Buckley-Geer}, {Burke}, {Carnero Rosell}, {Carrasco Kind}, {Carretero},
  {Castander}, {Cunha}, {Davis}, {De Vicente}, {Diehl}, {Doel}, {Eifler},
  {Flaugher}, {Fosalba}, {Frieman}, {Garc{\'\i}a-Bellido}, {Gruen}, {Gruendl},
  {Gutierrez}, {Hartley}, {Hollowood}, {Honscheid}, {James}, {Jeltema},
  {Kuropatkin}, {Li}, {Lima}, {Maia}, {Martini}, {Menanteau}, {Ogando},
  {Plazas}, {Roodman}, {Sanchez}, {Scarpine}, {Schindler}, {Smith},
  {Soares-Santos}, {Sobreira}, {Suchyta}, {Swanson}, {Tarle}, {Tucker}, \&
  {Walker}}]{Pirsiainen2018}
{Pursiainen}, M., {Childress}, M., {Smith}, M., {et~al.} 2018, \mnras, 481,
  894, \dodoi{10.1093/mnras/sty2309}

\bibitem[{{Rabinak} \& {Waxman}(2011)}]{Rabinak2011}
{Rabinak}, I., \& {Waxman}, E. 2011, \apj, 728, 63,
  \dodoi{10.1088/0004-637X/728/1/63}

\bibitem[{Rasmussen(2006)}]{Rasmussen06gaussianprocesses}
Rasmussen, C.~E. 2006, in Gaussian processes for machine learning (MIT Press)

\bibitem[{{Rest} {et~al.}(2018){Rest}, {Garnavich}, {Khatami}, {Kasen},
  {Tucker}, {Shaya}, {Olling}, {Mushotzky}, {Zenteno}, {Margheim},
  {Strampelli}, {James}, {Smith}, {F{\"o}rster}, \& {Villar}}]{Rest2018}
{Rest}, A., {Garnavich}, P.~M., {Khatami}, D., {et~al.} 2018, Nature Astronomy,
  2, 307, \dodoi{10.1038/s41550-018-0423-2}

\bibitem[{{Rosswog} {et~al.}(2018){Rosswog}, {Sollerman}, {Feindt}, {Goobar},
  {Korobkin}, {Wollaeger}, {Fremling}, \& {Kasliwal}}]{Rosswog2018}
{Rosswog}, S., {Sollerman}, J., {Feindt}, U., {et~al.} 2018, Astronomy and
  Astrophysics, 615, A132, \dodoi{10.1051/0004-6361/201732117}

\bibitem[{{Salbi} {et~al.}(2014){Salbi}, {Matzner}, {Ro}, \&
  {Levin}}]{Salbi2014}
{Salbi}, P., {Matzner}, C.~D., {Ro}, S., \& {Levin}, Y. 2014, \apj, 790, 71,
  \dodoi{10.1088/0004-637X/790/1/71}

\bibitem[{{Smartt} {et~al.}(2017){Smartt}, {Chen}, {Jerkstrand}, {Coughlin},
  {Kankare}, {Sim}, {Fraser}, {Inserra}, {Maguire}, {Chambers}, {Huber},
  {Kr{\"u}hler}, {Leloudas}, {Magee}, {Shingles}, {Smith}, {Young}, {Tonry},
  {Kotak}, {Gal-Yam}, {Lyman}, {Homan}, {Agliozzo}, {Anderson}, {Angus},
  {Ashall}, {Barbarino}, {Bauer}, {Berton}, {Botticella}, {Bulla}, {Bulger},
  {Cannizzaro}, {Cano}, {Cartier}, {Cikota}, {Clark}, {De Cia}, {Della Valle},
  {Denneau}, {Dennefeld}, {Dessart}, {Dimitriadis}, {Elias-Rosa}, {Firth},
  {Flewelling}, {Fl{\"o}rs}, {Franckowiak}, {Frohmaier}, {Galbany},
  {Gonz{\'a}lez-Gait{\'a}n}, {Greiner}, {Gromadzki}, {Guelbenzu},
  {Guti{\'e}rrez}, {Hamanowicz}, {Hanlon}, {Harmanen}, {Heintz}, {Heinze},
  {Hernandez}, {Hodgkin}, {Hook}, {Izzo}, {James}, {Jonker}, {Kerzendorf},
  {Klose}, {Kostrzewa-Rutkowska}, {Kowalski}, {Kromer}, {Kuncarayakti},
  {Lawrence}, {Lowe}, {Magnier}, {Manulis}, {Martin-Carrillo}, {Mattila},
  {McBrien}, {M{\"u}ller}, {Nordin}, {O'Neill}, {Onori}, {Palmerio},
  {Pastorello}, {Patat}, {Pignata}, {Podsiadlowski}, {Pumo}, {Prentice}, {Rau},
  {Razza}, {Rest}, {Reynolds}, {Roy}, {Ruiter}, {Rybicki}, {Salmon}, {Schady},
  {Schultz}, {Schweyer}, {Seitenzahl}, {Smith}, {Sollerman}, {Stalder},
  {Stubbs}, {Sullivan}, {Szegedi}, {Taddia}, {Taubenberger}, {Terreran}, {van
  Soelen}, {Vos}, {Wainscoat}, {Walton}, {Waters}, {Weiland}, {Willman},
  {Wiseman}, {Wright}, {Wyrzykowski}, \& {Yaron}}]{Smartt2017}
{Smartt}, S.~J., {Chen}, T.-W., {Jerkstrand}, A., {et~al.} 2017, \nat, 551, 75,
  \dodoi{10.1038/nature24303}

\bibitem[{{Street} {et~al.}(2018){Street}, {Lund}, {Khakpash}, {Donachie},
  {Dawson}, {Golovich}, {Wyrzykowski}, {Szkody}, {Naylor}, {Penny},
  {Rattenbury}, {Dall'Ora}, {Clarkson}, {Bennett}, {Pepper}, {Rabus},
  {Hundertmark}, {Tsapras}, {Di Stefano}, {Ridgway}, {Liu}, \&
  {Lin}}]{streetWP}
{Street}, R.~A., {Lund}, M.~B., {Khakpash}, S., {et~al.} 2018, arXiv e-prints,
  arXiv:1812.03137.
\newblock \doarXiv{1812.03137}

\bibitem[{{Stritzinger} {et~al.}(2018){Stritzinger}, {Shappee}, {Piro},
  {Ashall}, {Baron}, {Hoeflich}, {Holmbo}, {Holoien}, {Phillips}, {Burns},
  {Contreras}, {Morrell}, \& {Tucker}}]{Stritzinger2018}
{Stritzinger}, M.~D., {Shappee}, B.~J., {Piro}, A.~L., {et~al.} 2018, \apj,
  864, L35, \dodoi{10.3847/2041-8213/aadd46}

\bibitem[{{Tanvir} {et~al.}(2013){Tanvir}, {Levan}, {Fruchter}, {Hjorth},
  {Hounsell}, {Wiersema}, \& {Tunnicliffe}}]{Tanvir13}
{Tanvir}, N.~R., {Levan}, A.~J., {Fruchter}, A.~S., {et~al.} 2013, \nat, 500,
  547, \dodoi{10.1038/nature12505}

\bibitem[{{Tanvir} {et~al.}(2017){Tanvir}, {Levan},
  {Gonz{\'a}lez-Fern{\'a}ndez}, {Korobkin}, {Mandel}, {Rosswog}, {Hjorth},
  {D'Avanzo}, {Fruchter}, {Fryer}, {Kangas}, {Milvang-Jensen}, {Rosetti},
  {Steeghs}, {Wollaeger}, {Cano}, {Copperwheat}, {Covino}, {D'Elia}, {de Ugarte
  Postigo}, {Evans}, {Even}, {Fairhurst}, {Figuera Jaimes}, {Fontes}, {Fujii},
  {Fynbo}, {Gompertz}, {Greiner}, {Hodosan}, {Irwin}, {Jakobsson},
  {J{\o}rgensen}, {Kann}, {Lyman}, {Malesani}, {McMahon}, {Melandri},
  {O'Brien}, {Osborne}, {Palazzi}, {Perley}, {Pian}, {Piranomonte}, {Rabus},
  {Rol}, {Rowlinson}, {Schulze}, {Sutton}, {Th{\"o}ne}, {Ulaczyk}, {Watson},
  {Wiersema}, \& {Wijers}}]{Tanvir2017}
{Tanvir}, N.~R., {Levan}, A.~J., {Gonz{\'a}lez-Fern{\'a}ndez}, C., {et~al.}
  2017, \apjl, 848, L27, \dodoi{10.3847/2041-8213/aa90b6}

\bibitem[{{Tauris} {et~al.}(2015){Tauris}, {Langer}, \&
  {Podsiadlowski}}]{Tauris2015}
{Tauris}, T.~M., {Langer}, N., \& {Podsiadlowski}, P. 2015, \mnras, 451, 2123,
  \dodoi{10.1093/mnras/stv990}

\bibitem[{{The PLAsTiCC team} {et~al.}(2018){The PLAsTiCC team}, {Allam},
  {Bahmanyar}, {Biswas}, {Dai}, {Galbany}, {Hlo{\v{z}}ek}, {Ishida}, {Jha},
  {Jones}, {Kessler}, {Lochner}, {Mahabal}, {Malz}, {Mandel},
  {Mart{\'\i}nez-Galarza}, {McEwen}, {Muthukrishna}, {Narayan}, {Peiris},
  {Peters}, {Ponder}, {Setzer}, {The LSST Dark Energy Science Collaboration},
  {LSST Transients}, \& {Variable Stars Science Collaboration}}]{plasticc}
{The PLAsTiCC team}, {Allam}, Tarek, J., {Bahmanyar}, A., {et~al.} 2018, arXiv
  e-prints, arXiv:1810.00001.
\newblock \doarXiv{1810.00001}

\bibitem[{{Totani} \& {Panaitescu}(2002)}]{Totani2002}
{Totani}, T., \& {Panaitescu}, A. 2002, \apj, 576, 120, \dodoi{10.1086/341738}

\bibitem[{{Villar} {et~al.}(2017){Villar}, {Guillochon}, {Berger}, {Metzger},
  {Cowperthwaite}, {Nicholl}, {Alexander}, {Blanchard}, {Chornock},
  {Eftekhari}, {Fong}, {Margutti}, \& {Williams}}]{Villar2017}
{Villar}, V.~A., {Guillochon}, J., {Berger}, E., {et~al.} 2017, \apj, 851, L21,
  \dodoi{10.3847/2041-8213/aa9c84}

\bibitem[{{Yoon} \& {Cantiello}(2010)}]{Yoon2010}
{Yoon}, S.-C., \& {Cantiello}, M. 2010, \apj, 717, L62,
  \dodoi{10.1088/2041-8205/717/1/L62}

\end{thebibliography}

\end{document}